\documentclass[%
 aip,
rsi,%
 amsmath,amssymb,
reprint,%
]{revtex4-1}

\usepackage{graphicx}
\usepackage{dcolumn}
\usepackage{bm}
\usepackage{color}

\begin{document}


\title[Simulation of semidilute polymer solutions in planar extensional via conformationally averaged Brownian noise]
{Simulation of semidilute polymer solutions in planar extensional via conformationally averaged Brownian noise}

\author{Charles D. Young}

\affiliation{ 
Department of Chemical and Biomolecular Engineering, University of Illinois at Urbana-Champaign, Urbana, Illinois 61801, USA
}%

\author{Charles E. Sing}
 \email{cesing@illinois.edu}
\affiliation{ 
Department of Chemical and Biomolecular Engineering, University of Illinois at Urbana-Champaign, Urbana, Illinois 61801, USA
}

\date{\today}

\begin{abstract}
The dynamics and rheology of semidilute polymer solutions in strong flows are of great practical relevance. Processing applications can in principle be designed utilizing the relationship between nonequilibrium polymer conformations and the material properties of the solution. However, the interplay between concentration, flow, hydrodynamic interactions (HI), and topological interactions which govern semidilute polymer dynamics are challenging to characterize. Brownian dynamics (BD) simulations are particularly valuable as a way to direct visualize how molecular interactions arise in these systems, and are quantitatively comparable to single-molecule experiments. However such simulations are often computationally intractable, and are limited by the need to calculate the correlated Brownian noise via decomposition of the diffusion tensor. Previously we have introduced an iterative conformational averaging (CA) method for BD simulations which bypasses these limitations by preaveraging the HI and Brownian noise in an iterative procedure. In this work, we generalize the CA method to flowing semidilute solutions by introducing a conformation dependent diffusion tensor and a strain dependent approximation to the conformationally averaged Brownian noise. We find that this approach nearly quantitatively reproduces both transient and steady state polymer dynamics and rheology while achieving an order of magnitude computational acceleration. We then utilize the CA method to investigate the concentration and flow rate dependence of polymer dynamics in planar extensional flows. Our results are consistent with previous experimental and simulation studies and provide a detailed view of broad conformational distributions in the semidilute regime. We observe interconversion between stretched and coiled states at steady state, which we conjecture occur due to the effect of concentration on the conformation dependent polymer drag. Additionally, we observe transient flow-induced intermolecular hooks in the startup of flow which lead to diverse and unique stretching pathways.
\end{abstract}

\maketitle

\section{\label{sec:Intro}Introduction}

Polymer solution dynamics and rheology are relevant to polymer applications due to the relationship between molecular conformations and the applied processing flows. \cite{ferry1980viscoelastic, bird1987dynamics, schroeder2018single} In particular, flow-driven polymer alignment can have a significant impact on materials properties. For example, strong flows induced during solution coating affect the conductivity of organic thin film transistors, \cite{diao2013solution} and electrospun polymer fibers exhibit a wide range of flow-driven morphologies depending on the polymer concentration and molecular weight. \cite{huang2003review} In these empirical studies, flow can in principle be tuned to engineer for desired properties; nevertheless, a {\it molecular} view of out-of-equilibrium polymer dynamics would inform how these processing methods could be designed.

Despite this practical importance, it remains a challenge to predict polymer dynamics in processing flows because the solutions are often {\it semidilute}, meaning that the solution concentration is sufficiently high that polymer coils significantly interact. Formally, this is quantified as having a polymer concentration greater than the overlap concentration $c^{*}=M/(N_A R_g^{3})$, where $M$ is the polymer molecular weight, $N_A$ is Avogadro's number, and $R_g$ is the dilute polymer radius of gyration. \cite{de1979scaling, rubinstein2003polymer, doi1988theory} Below the overlap concentration $c^*$, interpolymer interactions are negligible such that the solution properties can be determined from the conformations of a single, isolated polymer. Above $c^*$, polymers overlap and intermolecular interactions must be considered. Additionally, solutions are generally processed in strong flows which disturb the polymers from their equilibrium conformations. The effects of concentration and flow on conformational dynamics are coupled, complicating the development of a comprehensive understanding of semidilute out-of-equilibrium dynamics.

Polymer dynamics are generally studied in the context of the classical theories of Rouse \cite{rouse1953theory} and Zimm, \cite{zimm1956dynamics} which both predict the diffusive motion of {\it individual} polymer chains modeled as bead-spring chains. The Rouse model neglects excluded volume (EV) and solvent-mediated hydrodynamic interactions (HI) and is thus appropriate for unentangled melts, where intra- and intermolecular interactions are screened by surrounding polymers. \cite{doi1988theory, rubinstein2003polymer} The Zimm model incorporates HI by preaveraging over the equilibrium polymer conformation and is accurate in the dilute limit where HI is unscreened. \cite{doi1988theory, rubinstein2003polymer}

Progress in polymer solution dynamics has built on the Zimm model to focus on the effect of {\it dilute} chains in strong flows, where individual chains can undergo a coil-stretch transition. Pioneering work by de Gennes demonstrated how strong, linear flows (e.g. shear and extensional flows) can stretch a polymer away from its equilibrium coiled conformation. \cite{de1974coil} In extensional flows, which we focus on in this paper, this occurs at a critical Weissenberg number $Wi = 0.5$, \cite{larson1989coil} where $Wi = \dot{\epsilon} \tau$ is related to the strain rate $\dot{\epsilon}$ and the longest polymer relaxation time $\tau$ predicted by Zimm. \cite{zimm1956dynamics} This initial picture has been refined by increasingly sophisticated theories, providing insight into planar mixed flows, the role of conformation-dependent HI, and the effect of solvent quality. \cite{bird1987dynamics2, ottinger1996stochastic, ottinger1987generalized, ottinger1989gaussian, magda1988deformation, larson1989coil, prakash1997universal, prakash1999kinetic}

Single molecule experiments \cite{perkins1997single, smith1998response, schroeder2003observation} and Brownian dynamics (BD) simulations \cite{larson1999brownian, jendrejack2002stochastic, schroeder2003observation} confirmed theoretical predictions for the coil-stretch transition in extensional flow. These studies also revealed unexpected behavior in transient conformations. In particular, identical polymers displayed `molecular individualism' \cite{de1997molecular} in stretching behavior, with a subpopulation stretching significantly slower than the ensemble average due to a specific initial conformation. In shear flows, polymers were observed to undergo tumbling cycles between coiled and stretched conformations on periodic time scales. \cite{schroeder2005characteristic, schroeder2005dynamics, teixeira2007individualistic} Experimental and simulation studies have since revealed that even in the dilute limit, polymer dynamics show non-trivial dependence on chain architecture, \cite{mai2015topology, hsiao2016ring,  mai2018stretching, young2019ring} solvent quality, \cite{somani2010effect, sing2010globule, sing2011dynamics} and flow. \cite{schroeder2004effect, hsieh2004modeling, hsieh2005prediction, sing2011dynamics, young2019ring}

Despite extensive studies of dilute solutions, understanding of semidilute solution dynamics remains limited. At equilibrium, scaling theories combining results of Rouse and Zimm models provide predictions for concentration dependence. \cite{de1976dynamics, rubinstein2003polymer, jain2012dynamic} Kinetic theories can also be extended to the semidilute regime, although accurately including the effects of surrounding polymers quickly becomes intractable. \cite{edwards1974theory, edwards1984brownian} Due to the challenge of simultaneously including the effects of HI and correlated semidilute solution structure, we are only aware of one theory for semidilute polymer dynamics in extensional flows; Prabhakar et al. \cite{prabhakar2016influence} have proposed a microstructural constitutive model combining `blob' arguments for concentration dependent drag on coiled and stretched polymers. The model shows that as concentration increases from the dilute limit to the overlap concentration, stretched polymers effectively cause the solution to self-concentrate relative to equilibrium. Above the overlap concentration, stretching allows polymers to reduce intermolecular interactions relative to the equilibrium case, and the solution thus self-dilutes. Some results of this model are consistent with experimental measurements,\cite{prabhakar2016influence} yet it remains challenging to extend equilibrium concepts to constitutive models for flowing polymers. Such approaches have found partial success for semidilute nonlinear shear rheology, \cite{heo2008universal} but there remains a need for a more universal molecular understanding. \cite{prakash2019universal}

Due to these challenges, progress towards understanding flowing semidilute polymer dynamics has primarily been made by experiment and simulation. Early studies were made by flow birefrigence, which found overshoots in the transient response. \cite{fuller1981flow, ng1993concentration} The steady state polymer extension as calculated from birefrigence was observed to decrease with concentration, indicating that increased intermolecular interactions inhibited stretching. Semidilute polymer rheology has been investigated by filament stretching experiments, \cite{tirtaatmadja1993filament} which showed that extensional viscosity was an order of magnitude less than Batchelor's prediction for a suspension of elongated particles. \cite{batchelor1971stress} This suggests polymers were not completely stretched at steady state. \cite{james1995molecular} Clasen et al. have investigated semidilute dynamics by capillary thinning. \cite{clasen2006dilute} In this study, the authors found the concentration dependence of the polymer relaxation time to be much stronger than comparable shear rheology experiments. Additionally, they found that polymer relaxation in extensional flow is concentration dependent significantly below the overlap concentration. The authors proposed this occurs because polymers `self-concentrate' in flow due to stretching, inspiring the theoretical approach of Prabhakar et al. described above. \cite{prabhakar2016influence}


Several single molecule studies have also been performed on semidilute solutions, although primarily in shear flows. \cite{hur2001dynamics, babcock2000relating, teixeira2007individualistic, harasim2013direct, huber2014microscopic} A recent study by Hsiao et al. \cite{hsiao2017direct} visualized single polymers in semidilute solution under planar extensional flow using a hydrodynamic trap, which held a tagged polymer at the stagnation point for extended periods. They observed steady state and transient dynamics by step strain rate experiments, in which polymers at equilibrium were suddenly exposed to a constant step strain rate until high levels of accumulated strain $\epsilon = \dot{\epsilon} t = 10-15$ and then relaxed to equilibrium after flow cessation. The experiments revealed broad conformational distributions under the startup of flow and increased molecular individualism, with different stretching pathways as compared to dilute solutions. Hsiao et al. attributed these differences to enhanced intermolecular interactions and the formation of transient flow-induced entanglements.

Simulations are a natural complement to experimental progress because they allow for direct study of intermolecular interactions. Early efforts by Stoltz et al. \cite{stoltz2006concentration} performed coarse-grained BD simulations of semidilute $\lambda$-DNA solutions in shear and planar extensional flows. They suggested that concentration effects on steady state polymer stretch are largely eliminated by plotting versus the concentration dependent Weissenberg number $Wi_c = \dot{\epsilon} \tau_c$, where $\tau_c$ is the longest polymer relaxation time at the relevant concentration. However, the reduced shear and extensional viscosity continued to increase monotonically with concentration above $Wi_c = 0.5$. The authors explained this in terms of the enhanced solvent velocity perturbations in the semidilute case, which screened HI and thus increased solvent drag on the polymers. Samsal et al. \cite{sasmal2017parameter} have performed BD simulations of semidilute DNA solutions which quantitatively match the experimental results of Hsiao et al. via a successive fine-graining protocol. Steady state results were consistent with the findings of Stoltz et al.\cite{stoltz2006concentration} The concentration dependence of transient dynamics was also studied by the stretching of polymers in the startup of planar extensional flow and relaxation after flow cessation.\cite{sasmal2017parameter}

It is clear that while there have been significant advances towards understanding semidilute out of equilibrium dynamics, there remain many unresolved questions. In particular, an understanding of how intermolecular interactions drive concentration and flow rate dependence is necessary. Simulation studies are of crucial importance here, as it appears the diversity of molecular conformations and topological and hydrodynamic interactions in semidilute solutions cannot be explained by mean field approximations or observation of individual trajectories. This is further evident in non-linear architectures, where interaction of architecturally specific polymer features with a semidilute background of chains lead to unexpected dynamics. \cite{zhou2019effect} Simulations at the level of an individual Kuhn step which enforce chain crossing constraints have revealed unique dynamics in semidilute solutions and melts under shear flow. \cite{huang2010semidilute, huang2011tumbling, nafar2015individual} We expect these distinctions to be more significant in extensional flows, where experiments show that interactions affect rheological behavior even at concentrations well below $c^*$. \cite{clasen2006dilute}

While many of these questions can, in principle, be studied using molecular simulation, it remains a computational challenge to access relevant time and length scales while retaining sufficient molecular detail. In the case of Brownian dynamics (BD) simulations, solvent-mediated hydrodynamic interactions (HI) are required to accurately describe polymer dynamics. Hydrodynamic interactions are typically calculated using an Ewald sum, the computational cost of which scales with the number of particles $N$ in the system as $O(N^{2})$. \cite{jain2012optimization} Additionally, a square root of the diffusion tensor is required so that the correlated Brownian noise satisfies fluctuation-dissipation. Decomposition of the diffusion tensor scales as $O(N^{2})-O(N^{3})$. \cite{ermak1978brownian, fixman1986construction, ando2012krylov, geyer2009n} The simulation literature contains several algorithms for accelerating these calculations by numerical approximation. Particle mesh Ewald sums can reduce the scaling of calculating HI and decomposing the diffusion tensor to approximately $O(N \textrm{log}N)$, \cite{guckel1999large, banchio2003accelerated, sierou2001accelerated, liu2014large, saadat2015matrix} and in some cases the scaling is nearly linear. \cite{fiore2017rapid} However, the prefactor of computational expense in this techniques is large, and opportunity - and need - for further acceleration remains.

We have previously introduced an iterative conformational averaging (CA) method for BD simulations of dilute flowing polymer dynamics \cite{miao2017iterative} and semidilute equlibrium dynamics. \cite{young2018conformationally} In this method, the conformation-dependent diffusion tensor is replaced by a single diffusion tensor, averaged over the non-equilibrium polymer conformations.\cite{miao2017iterative,young2018conformationally} In this work, we extend the CA method to semidilute solutions in planar extensional flows. Notably, we no longer preaverage the intramolecular HI over the out-of-equilibrium polymer conformation, which is important for adequately sampling the broad conformational distributions in semidilute solutions. We instead calculate the HI exactly for nearby polymer beads within a cutoff distance $r_c$ where HI decays quickly. Outside this cutoff, we use a discrete approximation to the exact diffusion tensor that is pre-computed before simulation. Effectively, only the Brownian noise remains conformationally averaged in this approach, which retains the computational speedup of the equilibrium semidilute CA method.\cite{young2018conformationally}

In this paper, we use the semidilute, out-of-equilibrium CA method to perform detailed investigation of polymer dynamics in planar extensional flow. The remainder of the article is organized as follows: In Section \ref{sec:goveqn} we state the governing equations for BD simulations. In Section \ref{sec:CAmethod} we introduce the semidilute out-of-equilibrium conformational averaging method. We then verify the CA method in Section \ref{sec:Verification} by comparison of steady state and transient dynamics and rheology to BD simulations in which the HI and Brownian noise are calculated without approximation. Next we consider the concentration and flow rate dependence via steady state and transient fractional extension and extensional viscosity. We also present individual molecular trajectories and conformational distributions. In Section \ref{sec:hooking}, we make a preliminary investigation of transient flow-induced intermolecular hooks. Finally, we summarize our results in Section \ref{sec:Conclusions} and suggest areas of interest for further study.


\section{\label{sec:Simulation Method}Simulation Method}

\subsection{\label{sec:goveqn}Governing Equations}

We perform BD simulations of semidilute polymer solutions in a planar extensional flow. The simulations consists of $N_{c}$ polymers each with $N_{b}$ coarse-grained bead such that the total number of beads is $N=N_{b}N_{c}$. The position $\bm{r}_{i}$ of a bead $i$ is updated according to the Langevin equation
\begin{equation}
    \frac{d\tilde{\bm{r}}_{i}}{d\tilde{t}} = \tilde{\bm{\kappa}} \cdot \tilde{\bm{r}}_{i} -\sum_{j} \tilde{\textbf{D}}_{ij} \nabla_{\tilde{\bm{r}}_{j}}(\tilde{U}) + \tilde{\bm{\xi}}_{i}
\end{equation}
Tildes denote dimensionless quantities. Positions are normalized by the bead radius ($\tilde{\bm{r}}=\bm{r}/a$), energies are normalized by the thermal energy $k_{B}T$ ($\tilde{U}=U/(k_{B}T)$), times are normalized by the single-bead diffusion time ($\tilde{t}=t/\tau_{0}$, where $\tau_{0}=6\pi \eta_s a^{3}/(k_{B}T)$ and $\eta_s$ is the solvent viscosity), and the diffusion tensor is normalized by the drag coefficient of the spherical polymer beads ($\tilde{\textbf{D}}_{ij}= \textbf{D}_{ij}(6\pi\eta a/k_B T)$). Polymer beads experience a planar extensional flow via the $3N \times 3N$ block diagonal tensor $\tilde{\bm{\kappa}}$, which has $3 \times 3$ diagonal blocks given by the solvent velocity gradient tensor $(\nabla \tilde{\textbf{v}})^T$. For planar extensional flow,
\begin{equation}
    \nabla \tilde{\textbf{v}} = \begin{pmatrix} \tilde{\dot{\epsilon}} & 0 & 0 \\ 0 & -\tilde{\dot{\epsilon}} & 0 \\ 0 & 0 & 0 \end{pmatrix}
\end{equation}
where $\tilde{\dot{\epsilon}} = \dot{\epsilon} \tau_{0}$ is the dimensionless strain rate. Beads interact via a potential $\tilde{U} = \tilde{U}^{B} + \tilde{U}^{EV}$ consisting of bonded and excluded volume contributions. We use a finitely extensible non-linear elastic (FENE) spring force for connectivity
\begin{equation}
    \tilde{U}^{B} = -0.5 \tilde{k}_{s} \tilde{r}_{max}^{2} \textrm{ln} \left[1-\left( \frac{\tilde{r}_{ij}}{\tilde{r}_{max}} \right)^{2} \right]
\end{equation}
where $\tilde{k}_{s}=30 \tilde{u}/\tilde{\sigma}^{2}$ is the spring constant, $\tilde{u}=1.0$ gives the strength of EV interactions, and $\tilde{\sigma}=2$ is the diameter of a bead. The maximum extension of a spring is $\tilde{r}_{max}=1.5\tilde{\sigma}$, and $\tilde{r}_{ij}$ is the distance between two connected beads. Excluded volume interactions are modeled by a shifted, purely repulsive Lennard-Jones potential
\begin{equation}
    \tilde{U}^{EV} = 4\tilde{u} \left[ \left( \frac{\tilde{\sigma}}{\tilde{r}_{ij}} \right)^{12} - \left( \frac{\tilde{\sigma}}{\tilde{r}_{ij}} \right)^{6} + \frac{1}{4} \right ] \Theta(2^{1/6} \tilde{\sigma} - r)
\end{equation}
which yields chain statistics representative of a good solvent. We find $\nu \approx 0.59$ from the scaling relation $\tau_{Z}\sim N^{3\nu}$ and relaxation time data from equilibrium single chain simulations, in agreement with the result for a polymer in good solvent. This model has been widely utilized to study polymer dynamics in solution and melt \cite{kremer1990dynamics} and has been shown to prevent chain crossings in simulations of entangled melts in extensional flow. \cite{kroger1997polymer, xu2018molecular, o2018relating}

Solvent-mediated HI and Stokes drag and included via the diffusion tensor, given here by the Rotne-Prager Yamakawa (RPY) tensor, \cite{rotne1969variational, yamakawa1970transport}
\begin{equation}
	\label{RPY}
    \tilde{\textbf{D}}_{ij} = \small \begin{cases}
    	\bm{I}, & i=j\\
        \frac{3}{4\tilde{r}_{ij}}\left[\left(1+\frac{2}{3{\tilde{r}_{ij}}^2}\right)\bm{I}+\left(1-\frac{2}{{\tilde{r}_{ij}}^2}\right)\bm{\hat{r}}_{ij}\bm{\hat{r}}_{ij}\right], & i\neq j, {\tilde{r}_{ij}}\geq 2\\
        \left(1-\frac{9{\tilde{r}_{ij}}}{32}\right)\bm{I}+\frac{3{\tilde{r}_{ij}}}{32}\hat{\bm{r}}_{ij}\hat{\bm{r}}_{ij}, & i\neq j, {\tilde{r}_{ij}}\leq 2\\
    \end{cases}
\end{equation}
$\hat{\bm{r}}_{ij} = \tilde{\bm{r}}_{ij}/r_{ij}$ is a unit vector in the direction of $\tilde{\bm{r}}_{ij}=\tilde{\bm{r}}_{j}-\tilde{\bm{r}}_{i}$ and $\bm{I}$ is the identity matrix. The average and first moment of the Brownian noise $\tilde{\bm{\xi}}_{i}$ are given by the fluctuation-dissipation theorem as $\langle \tilde{\bm{\xi}}_{i}(t) \rangle = 0$ and $\langle \tilde{\bm{\xi}}_{i}(t) \tilde{\bm{\xi}}_{j}(t') \rangle =2\tilde{\textbf{D}}_{ij}\delta(t-t')$ respectively. Simulation implementation requires the decomposition of the diffusion tensor as $\tilde{\textbf{D}} = \textbf{BB}^{T}$ so that the Brownian noise can be computed via $\tilde{\bm{\xi}}_{i}=\sqrt{2}\textbf{B}_{ij} \bm{f}_{j}$, where $\bm{f}_{j}$ is a Gaussian random variable with mean 0 and variance 1.

Polymers are simulated in an initially cubic simulation cell of size $\tilde{V} = \tilde{l}^{3}$. The cell size is determined by $\tilde{V}=N/\tilde{c}$, where $\tilde{c}$ is the polymer concentration. We set the concentration via the normalized concentration $\tilde{c}/\tilde{c}^{*}$, where $\tilde{c}^{*}=N_{b}/(4/3 \pi \langle \tilde{R}_{g0} \rangle ^{3})$ is the overlap concentration. The polymer conformations are initialized randomly and relaxed by an equilibrium run of approximately $10 \tau_{R}$ (where $\tau_{R}$ is the Rouse relaxation time in dilute solution) in which HI are neglected. This is followed by a production run including flow and HI. Flow is implemented using Kraynik-Reinelt boundary conditions (KRBCs) \cite{kraynik1992extensional} such that the simulation cell deforms consistently with the applied flow. We follow the implementation of Todd and Daivis,\cite{todd1998nonequilibrium} which allows for unrestricted strain accumulation. Hydrodynamics are accounted for using an Ewald sum, \cite{beenakker1986ewald, jain2012optimization} which overcomes the slow convergence of the RPY tensor by splitting the sum into exponentially decaying real space and reciprocal space parts. The full details are available in the appendix, but the implementation is similar to the equilibrium case except for the basis vectors, which vary as the cell deforms. Once the specified strain $\epsilon_{total}$ is reached, flow is ceased and the polymers relax to their equilibrium conformations. During cessation, cell deformation is halted and the cell remains in the configuration at the cessation time $t_{cess}=\epsilon_{total}/(\dot{\epsilon}dt)$. The simulation is advanced by explicit Euler integration of the Langevin equation using a time step of $dt=2-7 \times 10^{-4} \tau_{0}$, where higher concentration and strain rate simulations require smaller time steps.

\subsection{\label{sec:CAmethod}Iterative conformational averaging method}

BD simulations are typically computationally limited by the decomposition of the diffusion tensor to calculate the correlated Brownian noise.\cite{ermak1978brownian} In semidilute simulations, evaluation of the Ewald sum HI is also a significant expense.\cite{beenakker1986ewald} Progress towards accelerating BD simulations has mostly come from mathematical approximations discussed in the Introduction. Alternatively, we have recently made a physical approximation which assumes that HI can be conformationally averaged (CA) over the polymer conformation and the structure of the surround solution.\cite{young2018conformationally} An averaged form of the Brownian noise can also be constructed to satisfy the fluctuation-dissipation theorem. This approach effectively avoids the expense of the Ewald sum and diffusion tensor decomposition which must typically be computed $O(10^{6}$ times over the course of a simulation. Thus the Langevin equation takes the form
\begin{equation}
    \label{eqn:LangevinCA}
    \frac{d\tilde{\bm{r}}_{i}^{(w)}}{d\tilde{t}} = \tilde{\bm{\kappa}} \cdot \tilde{\bm{r}}_{i}^{(w)} -\sum_{j} \langle \tilde{\textbf{D}}_{ij} \rangle ^{(w)} \nabla_{\tilde{\bm{r}}_{j}}(\tilde{U}) +     \tilde{\bm{\xi}}_{i}^{(w)}
\end{equation}
Here we have introduced a CA approximation to the diffusion tensor, $\langle \tilde{\textbf{D}}_{ij} \rangle ^{(w)}$. Fluctuation-dissipation is maintained by a CA Brownian noise $\tilde{\bm{\xi}}_{i}^{(w)}$, which satisfies $\langle \tilde{\bm{\xi}}_{i}^{(w)}(t) \tilde{\bm{\xi}}_{j}^{(w)}(t') \rangle = 2 \langle \tilde{\textbf{D}}_{ij} \rangle ^{(w)} \delta(t-t')$. The challenge in this case is finding the correct average consistent with the polymer conformational distribution. Previously we have introduced an iterative procedure to self-consistently determine the conformationally averaged HI and Brownian noise for single chain and semidilute systems. In this approach, the superscript $(w)$ indicates the iteration number. For a complete description of the single chain and semidilute equilibrium methods, we refer to our previous publications. Here we briefly outline the method
\begin{enumerate}
	\item Begin with a guess for the averaged diffusion tensor. For simplicity, we typically use the freely draining (FD) case, $\langle \tilde{\textbf{D}}_{ij} \rangle ^{(0)} = \delta_{ij} \bm{I}$. More informed guesses can also be made, for example using Zimm theory.
	\item Perform a BD simulation using the averaged diffusion tensor to determine the average polymer conformational distribution.
	\item Use the simulation results to calculate the CA diffusion tensor and Brownian noise for this iteration. Assuming steady state, the phase space average can be replaced with a time average where the conformational probability is discretely sampled via the BD simulation.
	\item  Repeat steps 2 and 3 with increasing values of $w=1,2,3,...$ until converged to a self-consistent diffusion tensor and Brownian noise.
\end{enumerate} 
Each iteration is run for $15 \tau_{R}^{(w)}$, where $\tau_{R}^{(w)}$ is the longest relaxation time of a single chain for iteration $(w)$. We find this provides sufficient sampling of conformational space. In the first iteration using freely draining dynamics, $\tau_R^{(w)}$ is the Rouse relaxation time. For subsequent iterations including HI, this is the Zimm relaxation time. 
\subsection{CA for semidilute out of equilibrium solutions}
Previous work by the authors has shown that the CA method quantitatively matches results from traditional BD simulations for dilute linear and ring polymer solutions in equilibrium and under steady state flow, with the exception of strong shear flows that exhibit large conformational fluctuations. \cite{miao2017iterative} We have also extended the method to semidilute equilibrium systems by utilizing a discrete approximation to the RPY tensor for intermolecular HI to account for fluctuations as chain center of masses diffusive relative to each other over time. \cite{young2018conformationally} We have also constructed the Brownian noise to be consistent with these fluctuations using a modified version of the truncated expansion ansatz (TEA). \cite{geyer2009n} The semidilute CA method then quantitatively reproduces diffusion constant results of traditional BD simulations using the TEA, and also reproduces static properties and zero-shear viscosity from simulations using the Krylov subspace decomposition up to small quantitative differences. Out out of equilibrium simulations are more sensitive to approximations, however, \cite{saadat2014computationally} so we must modify our equilibrium semidilute method.
\subsubsection{Grid space diffusion tensor approximation}

\begin{figure*}[htb]
	\includegraphics[width=\textwidth]{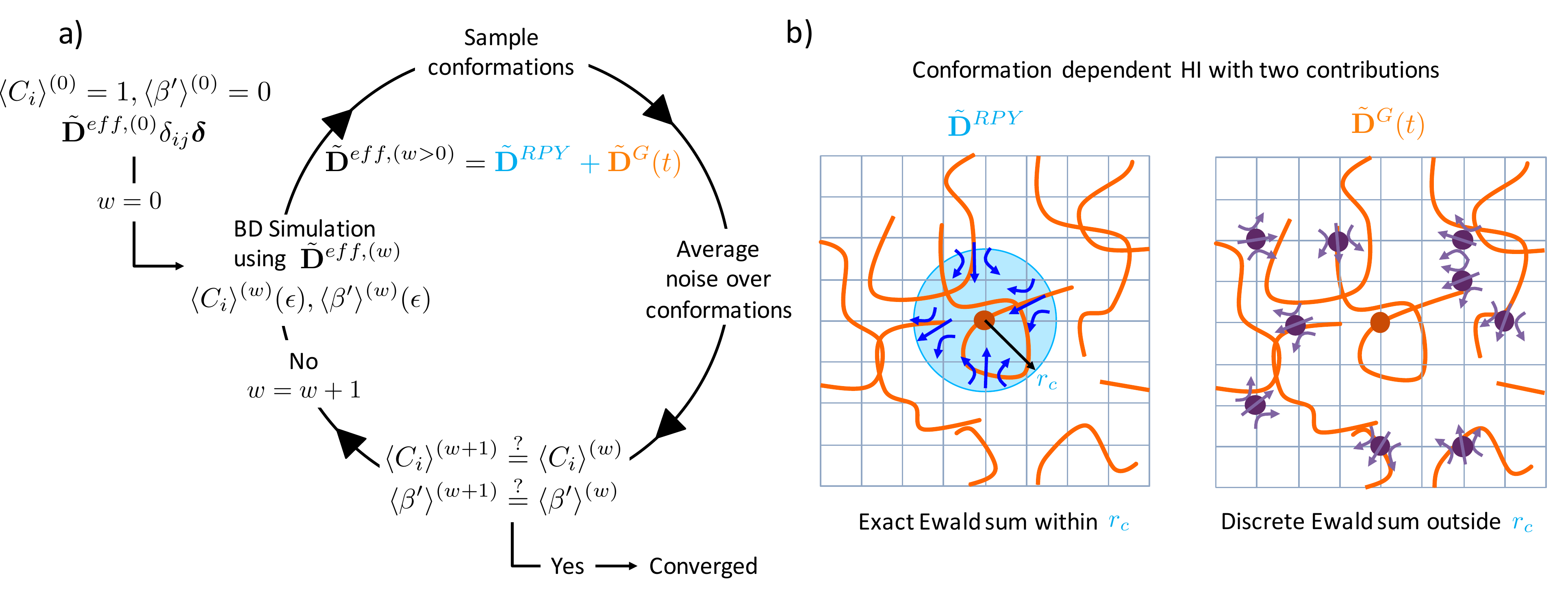}
    \caption{a) Schematic describing the out-of-equilibrium iterative conformational averaging method. For the first iteration $w=0$, a freely draining diffusion tensor and Brownian noise are used to a run a BD simulation under planar extensional flow. Throughout the simulation, polymer conformations are sampled to determined to average Brownian noise as a function of accumulated strain via the TEA parameters $\langle C_i \rangle^{(w=1)}(\epsilon), \langle \beta' \rangle^{(w=1)} (\epsilon)$. The averaged noise is then used to perform another simulation in which the diffusion tensor is given by the exact and grid space portions. b) For iterations $w>0$, we construct the diffusion tensor by an exact Ewald sum $\tilde{\textbf{D}}^{RPY}$ for beads with a relative separation $r_{ij}$ less than a cutoff radius $r_c$. Hydrodynamic interactions for beads with $r_{ij} > r_c$ are determined by a pre-computed discrete approximation to the exact Ewald sum, $\tilde{\textbf{D}}^{G}(t)$, which varies with the deformation of the cell to account for the change in the reciprocal space contribution. }
    \label{fig:schematic}
\end{figure*}

The first change we make is to construct the diffusion tensor of two parts schematically depicted in Figure \ref{fig:schematic}b i) an exact calculation from a full Ewald sum for neighboring beads within $r_{c}=10a$ ii) a discrete grid space approximation to the RPY tensor for beads outside $r_{c}$. We find that the conformational average for intramolecular HI previously used is not sufficient for simulations under planar extension, as conformational fluctuations are large and even slight deviations in the diffusion tensor can lead to significant errors in solution dynamics. Furthermore, we find that hydrodynamic coupling to neighbors within a short cutoff radius $r_{c}$ is strong enough to require a full Ewald sum calculation. Outside of this cutoff, the HI is sufficiently slowly varying that we can use a discrete representation of the RPY tensor in which the HI at grid displacements has been calculated ahead of time and stored in memory. This significantly reduces computational costs relative to performing the full Ewald sum for all pairs, as most pair displacements are greater than the cutoff, while retaining the accuracy required to reproduce the results of the full simulation. We generally find $r_{c}=10a$ to be sufficient, although for high density and strain rate simulations, a larger cutoff may be required. Thus the effective diffusion tensor can be written as
\begin{equation}
    \tilde{\textbf{D}}_{ij}^{eff} = \tilde{\textbf{D}}_{ij}^{RPY} \Theta(\tilde{r}_{c} - \tilde{r}_{ij}) + \tilde{\textbf{D}}_{ij}^{G}(t) \Theta(\tilde{r}_{ij} - \tilde{r}_{c})
\end{equation}
where $\Theta(x)$ is the Heaviside function, the $RPY$ superscript denotes the full Ewald sum, and $G$ the grid space. Here we have forgone the iteration counter $(w)$ and average brackets as this effective HI does not change from one iteration to the next and is not averaged over polymer conformations. For bead pairs outside $r_{c}$, the displacement $\tilde{\bm{r}}_{ij}$ is rounded to the nearest grid point $\Delta \tilde{\bm{r}}_{ij} = (\Delta \tilde{x}_{ij}, \Delta \tilde{y}_{ij}, \Delta \tilde{z}_{ij})$, and the effective HI for this pair is given by $\tilde{\textbf{D}}_{ij}^{G} = \tilde{\textbf{D}}^{RPY}(\Delta\tilde{\bm{r}}_{ij})$. 

For simulation results presented here, we use a grid spacing $d_{g}=1a$ between points for all but the highest concentration and strain rate condition. In the case of $N_{b}=100$, $c/c^{*}=3.0$, $Wi_0=3.0$ (where $Wi_0=\tilde{\dot{\epsilon}} \tau_{Z0}$ is the dimensionless strain rate normalized by the dilute Zimm relaxation time), we find that a more refined spacing of $d_{g}=0.5a$ is required for an accurate simulation. In this case, hydrodynamic coupling is strong leading to collective motions which can be sensitive to coarse graining of the HI. We use a uniform grid which is identical in $x$, $y$, and $z$. However, it is possible to use a spatially-varying grid size to overcome prohibitive memory requirements in the case of large simulation cells, designed so that nearby (i.e. stronger HI) interactions are determined more accurately than distant interactions.

A challenge that arises with the grid space HI for out of equilibrium simulations is that the basis vectors of the simulation cell change as the simulation progresses. While the real space part of the Ewald sum HI remains correct under application of periodic boundary conditions, the reciprocal space part varies as described in Appendix \ref{sec:appendix}. Therefore, we must pre-compute the Ewald sum at the specified grid displacements for various levels of cell deformation as encountered in simulation. Because KRBCs are periodic in time, these values can be used repeatedly over the course of the simulation. We find that $n_{g}=8$ samples are sufficient to describe the variation in the Ewald sum HI over the course of the cell deformation period. An alternative approach may be required for more flow types without time periodic boundary conditions, such as uniaxial extension. \cite{dobson2014periodic, hunt2016periodic} Additionally, as the simulation cell deforms from its initial state and is no longer cubic, the range of displacements stored in the grid space HI must increase accordingly. Thus the total number of grid points is given by $M = M_{x}M_{y}M_{z}$, where $M_{i}=\Delta i_{max}/d_{g}+1$, and $\Delta i_{max}$ is the largest displacement in the $i$ direction encountered over the course of a KRBC period. Then the total size of the array storing grid space displacement HI is given by $\tilde{\textbf{D}}^{G}(n_{g},9M)$. For most simulations presented here (up to $N=12,000$), we find the grid space array requires only ca. 1-2GB of memory.
\subsubsection{Strain dependent Brownian noise}
Having constructed the diffusion tensor, we also need to determine the correlated Brownian noise. Previously, we have used the TEA, which gives the noise vector as
\begin{equation}
	\label{eqn:TEA1}
    	\tilde{\xi}_{l} = \tilde{\textrm{D}}_{ll} C_{l} \sum_{m=1}^{3N} \beta' \frac{\tilde{\textrm{D}}_{lm}}{\tilde{\textrm{D}}_{ll}} f_{m}
\end{equation}
This notation follows that of Geyer and Winter for a single component of the size $3N$ noise vector, where the sum $m$ is over entries in a single row $l$ of the mobility matrix and does not follow index notation. This contrasts with the bead index notation used above, in which $\tilde{\bm{\xi}}_{i}$ gives the three components of the noise vector for a bead $i$. In the TEA, there are $3N$ coefficients $C_{l}$ which ensure each bead recovers the correct self-mobility and a weighting factor $\beta'$ that is chosen to approximate the diffusion tensor decomposition and satisfy fluctuation-dissipation via $\textrm{cov}(\tilde{\bm{\xi}},\tilde{\bm{\xi}}) \approx \tilde{\textbf{D}}$. Diagonal entries of the sum ($l=m$) use the weighting factor $\beta'=1$. All off diagonal entries use the same factor, given by
\begin{equation}
	\label{TEA3}
    \beta' = \frac{1-\sqrt{1-3N(\varepsilon^{2}-\varepsilon)}}{3N(\epsilon^{2}-\varepsilon)}
\end{equation}
where $\varepsilon$ is an average over the off-diagonal entries of the diffusion tensor
\begin{equation}
	\label{TEA4}
    \varepsilon = \frac{1}{(3N)^{2}}\sum_{l} \sum_{m \ne l} \frac{\tilde{\textrm{D}}_{lm}}{\tilde{\textrm{D}}_{ll}}
\end{equation}
The coefficients are given by
\begin{equation}
	\label{TEA2}
    C_{l} = \sqrt{\frac{1}{1+\beta'^{2} \sum_{l} \sum_{m \ne l} \frac{\tilde{\textrm{D}}_{lm}^{2}}{\tilde{\textrm{D}}_{ll}\tilde{\textrm{D}}_{mm}}}}
\end{equation}
Details on the derivation of the TEA can be found in the original paper from Geyer and Winter. \cite{geyer2009n} In the semidilute equilibrium case, the TEA parameters $\beta'$ and $C_{l}$ were conformationally averaged by a time average. In the out of equilibrium case, the conformation and thus the TEA parameters vary with the accumulated strain $\epsilon = \dot{\epsilon} t$. As a result, we must collect different averages as the fluid is deformed by flow. We choose discrete values of strain to bin the averages, yielding the strain-dependent average TEA parameters for a given iteration
\begin{equation}
    \label{eqn:BTEA}
    \langle \beta' \rangle^{(w)}(\epsilon_{o}) = \frac{1}{T}\sum_{\epsilon_o t_{\epsilon_{bin}}}^{(\epsilon_o+1) t_{\epsilon_{bin}}}{\beta'^{(w)}(t)} \\
\end{equation}
\begin{equation}
    \label{eqn:CTEA}
    \langle C_{l} \rangle^{(w)}(\epsilon_{o}) = \frac{1}{T}\sum_{\epsilon_o t_{\epsilon_{bin}}}^{(\epsilon_o+1) t_{\epsilon_{bin}}}{C^{(w)}_{l}(t)}
\end{equation}
where $\epsilon_{o}=\textrm{floor}(\epsilon/\epsilon_{bin})$ is the strain bin index, $t_{\epsilon_{bin}}$ is the number of time steps in a strain bin, and $T$ is the number of samples per bin. After the system reaches steady state, the TEA parameters can again be averaged to a single bin as in the equilibrium case. We define an equilibrium strain $\epsilon_{ss}$ and a number of strain bins $n_{\epsilon}$ to give the size of each strain bin $\epsilon_{bin}=\epsilon_{ss}/n_{\epsilon}$. The number of time steps per bin is $t_{\epsilon_{bin}} = \epsilon_{bin}/(\dot{\epsilon} dt)$. After flow cessation, the conformational distribution becomes time dependent again, and we resume dynamic averaging as in the startup transient case except over discrete time bins rather than strain bins. The flow cessation time is $t_{cess}$, and the system is allowed to relax for $t_{relax}=t_{max}-t_{cess}$. We set the number of time bins $n_{t}$, and the number of time steps per bin is $t_{t_{bin}}=t_{relax}/n_{t}$. The sums in Eqns. \ref{eqn:BTEA} and \ref{eqn:CTEA} are modified accordingly. For simulations presented here, we find $\epsilon_{ss}=10$, $n_{\epsilon}=20$, and $T=20$ provide sufficient resolution of the strain dependence to nearly quantitative match the transient response of the full BD simulation (Section \ref{sec:Verification}) while maintaining a low computational cost of sampling the Brownian noise. During relaxation, we use $t_{relax}=10\tau_{Z}$, $n_{t}=20$, and $T=20$. The Brownain noise can then be expressed as
\begin{equation}
	\label{eqn:TEACA}
	\begin{aligned}
    \tilde{\xi}_{l}^{(w)}(\tilde{t},\epsilon_{o}) ={} & \langle \tilde{\textrm{D}}_{ll} \rangle ^{(w)}(\tilde{t}) \langle C_{l} \rangle ^{(w)} (\epsilon_{o})  \langle \beta' \rangle ^{(w)} (\epsilon_{o}) \times \\ 
     & \sum_{m=1}^{3N} \frac{\langle \tilde{\textrm{D}}_{lm} \rangle ^{(w)}(\tilde{t})}{\langle \tilde{\textrm{D}}_{ll} \rangle ^{(w)}(\tilde{t})} f_{m}(\tilde{t})
    \end{aligned}
\end{equation}
and the Langevin equation as
\begin{equation}
    \label{eqn:LangevinCA2}
    \frac{d\tilde{\bm{r}}_{i}^{(w)}}{d\tilde{t}} = \tilde{\bm{\kappa}} \cdot \tilde{\bm{r}}_{i}^{(w)} -\sum_{j} \tilde{\textbf{D}}_{ij}^{eff} \nabla_{\tilde{\bm{r}}_{j}}(\tilde{U}) +     \tilde{\bm{\xi}}_{i}^{(w)}(\epsilon_o)
\end{equation}
\subsection{Computational cost}
Previously we have investigated the computational time scaling of the CA method with the number of beads $N$ at equilibrium.\cite{young2018conformationally} We found an order of magnitude speedup compared to traditional BD simulations using the Krylov subspace and TEA for decomposition of the diffusion tensor.\cite{young2018conformationally} In the non-equilibrium case, the most expensive operations of the CA method, namely the construction of the diffusion tensor and the matrix vector products (Eqs. \ref{eqn:TEACA} and \ref{eqn:LangevinCA2} second term), are comparable to the equilibrium case. While there are considerable qualitative differences in the details to improve accuracy, these come at only a small computational cost. The non-equilibrium CA method generally requires more memory for storing the grid space HI because of the need to pre-compute $n_{g}$ grids for the various states of cell deformation. For the simulations presented, however, we find the memory access speed is similar to the equilibrium case, and the memory requirements are not prohibitive. Thus we refer to our earlier work for a more detailed discussion of computational cost.\cite{young2018conformationally}

\section{\label{sec:Results}Results}

\begin{figure*}[htb]
	\includegraphics[width=\textwidth]{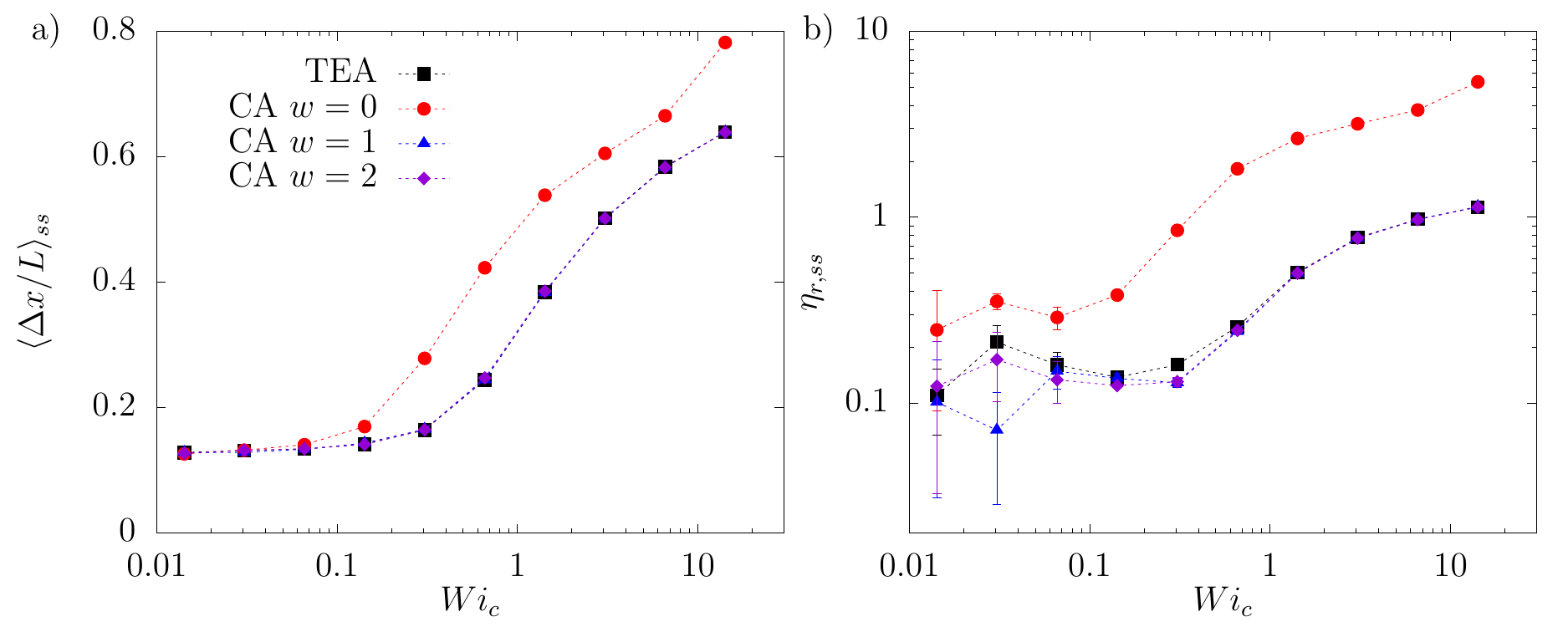}
    \caption{Steady state (a) fractional extension (b) reduced extensional viscosity as a function of flow strength $Wi_c$ for the TEA method (black squares), the first iteration $w=0$ (red circles), second iteration $w=1$ (blue triangles), and third iteration $w=2$ (purple diamonds) of the CA method. $N_{b}=50, N_{c}=40, c/c^{*}=0.3$.}
    \label{fig:verification_ss}
\end{figure*}

\subsection{\label{sec:Verification}Verification of the CA method}
We first verify the steady state dynamics and rheology of the CA method by comparison to BD simulations in which the Ewald sum is computed without approximation and the TEA without conformational averaging is used to compute to Brownian noise, which we refer to here as `traditional' or `TEA' BD simulations. While there are inherent approximations in the TEA,\cite{geyer2009n} we have previously shown that these lead to only small errors in the static polymer size and do not alter dynamic properties relative to simulations using the Krylov subspace. \cite{young2018conformationally} We expect the method to match best in this case as shown by theory and our previous simulations,\cite{miao2017iterative,young2018conformationally} which focus on steady state dynamics. We consider the ensemble average fractional extension projected along the axis of flow extension, $ \Delta x_{f}/L = (\textrm{max}(\{x_i\}) - \textrm{min}(\{x_i\}))/L$, where $L$ is the contour length. We also calculate the reduced extensional viscosity, in which we have normalized by the monomer concentration to account for the linear concentration dependence and used $c^*$ as the reference concentration
\begin{equation}
    \eta_r = \frac{\eta_p c^{*}}{\eta_s c}
\end{equation}
where $\eta_p$ is the polymer contribution to the extensional viscosity
\begin{equation}
    \eta_{p}=-\frac{\tau_{p,xx}-\tau_{p,yy}}{\dot{\epsilon}}
\end{equation}
The polymer contribution to the stress tensor is determined by the Kirkwood formula\cite{doi1988theory}
\begin{equation}
    \tau_{p,\alpha \beta} = \frac{1}{V} \sum_{i}^{N} \sum_{j>i}^{N} r_{ij,\alpha} F_{ij,\beta}
\end{equation}
where $F_{ij,\beta}$ is the conservative force between particles $i$ and $j$ in the $\beta$ direction. For this set of verification simulations, we model chains of length $N_{b}=50$ at a concentration $c/c^{*}=0.3$ with $N_{c}=40$ chains per simulation at strain rate $\dot{\epsilon}=1 \times 10^{-4} - 0.1$, for which traditional BD simulations are tractable.

\subsubsection{Steady state extension and viscosity}

For all simulations, the chains undergo a transition from a coiled conformation at low strain rate, where the polymer can relax strain more quickly than it is accumulated, to a stretched conformation in the extension direction $x$ at high strain rate (Fig. \ref{fig:verification_ss}a). The first iteration of the CA method ($w=0$) does not include HI, meaning the polymers are unshielded from flow. As a result, polymers stretch at a lower strain rate $\dot{\epsilon}$ relative to BD simulations with HI. Here we have plotted results from both FD and HI simulations against the flow strength $Wi_c = \dot{\epsilon} \tau_c$, where $\tau_c$ is the longest polymer relaxation time obtained from the HI simulations. We find the relaxation time by measuring the extension after flow cessation at steady state for the highest flow rate simulation, $Wi_c=17.4$. We then fit an exponential to the relaxation response for $\Delta x/L<0.2$ of the form $\langle \Delta x^2 \rangle / L^2 = (\Delta x_0^2 - \Delta x_\infty^2)e^{-t/\tau_c} + \Delta x_\infty^2$,\cite{hsiao2017direct} where $\Delta x_0=0.2$ is the value from which the fit starts, and $\Delta x_\infty$ is the constant long time limit extension.

Hydrodynamics and conformationally averaged Brownian noise are included in the second iteration $w=1$ of the CA method. In this case, the steady state fractional extension quantitatively matches the results of the traditional BD simulation. Unlike our previous work, where a further iteration $w=2$ was sometimes required to quantitatively reproduce steady state dynamics,\cite{miao2017iterative,young2018conformationally} in this case we find almost no change between the second and third iterations. This is explained by modifications of the diffusion tensor, which in our previous work varied from one iteration to the next with the polymer conformation. In the current implementation, the grid space HI is pre-computed once before simulation and the same values are used for all iterations. Therefore, the only variation in the Langevin equation from iteration to the next is in the conformationally averaged TEA parameters for the Brownian noise. These differences are generally small, and in the steady state case are negligible. Below we show that there are small quantitative changes to the transient extension and conformational distributions associated with the TEA parameters. Beyond Section \ref{sec:Verification}, we perform only two iterations, which provide nearly quantitatively accurate dynamics.

We also compare the extensional viscosity in Fig \ref{fig:verification_ss}b. The results are qualitatively similar to fractional extension because the spring force dominates the conservative force for a stretched polymer. The qualitative differences between simulations with and without HI are highlighted by the fact that at the same fractional extension, the simulation without HI shows a higher extensional viscosity despite a lower strain rate. In the FD case, all beads are unshielded from flow and solvent drag on the chain is large. The finitely extensible bonds in the center of the chain are thus under high tension. When HI is included and the chains become shielded from flow, however, the polymer is under less tension even at the same fractional extension. The extensional viscosity of the $w=1$ simulation also quantitatively matches the full BD simulation. 
\begin{figure*}[htb]
	\includegraphics[width=\textwidth]{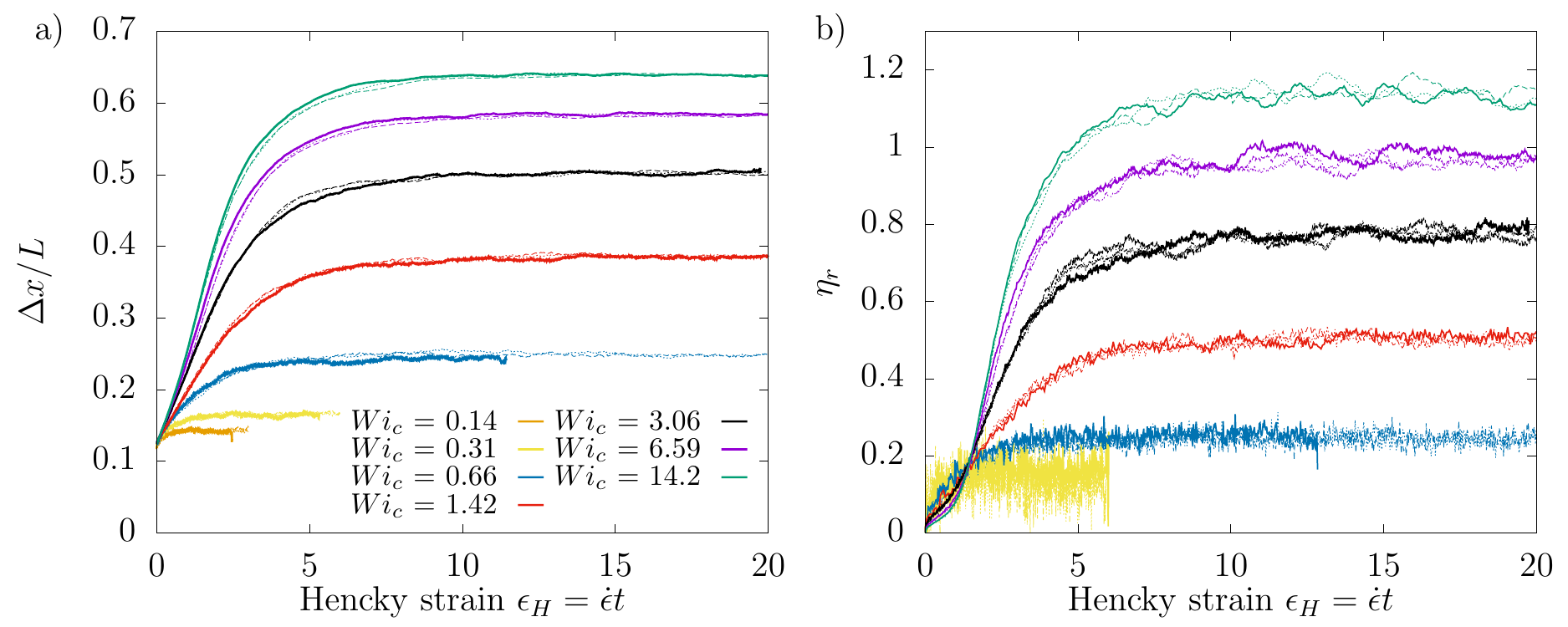}
    \caption{Transient (a) fractional extension (b) extensional viscosity as a function of accumulated strain at various strain rates for the TEA method (solid lines), the CA method $w=1$ (dashed lines), and the CA method $w=2$ (dotted lines). }
    \label{fig:verification_trans}
\end{figure*}

In the remainder of this work, we generally do not discuss the FD results from the $w=0$ iteration, and instead include the FD results in the Supplementary Information for comparison.

\subsubsection{Transient extension and viscosity}

We evaluate the accuracy of the transient dynamics and rheology of the second iteration and third iterations $w=1,2$ of the CA method in Figure~\ref{fig:verification_trans}. The fractional extension, in Figure~\ref{fig:verification_trans}a, increases from the equilibrium value of $\Delta x/L \approx 0.1$ to the steady state value shown in Fig \ref{fig:verification_ss} after an accumulated strain of approximately $\epsilon=10$. At lower strain rates, the polymer has more time to relax its conformation as strain is accumulated, so steady state is reached more quickly. We note minor quantitative differences between the second iteration $w=1$ of the CA method and the full BD simulation leading up to steady state, seen in the difference between the dotted and dashed line in Figure~\ref{fig:verification_trans}a, but there is no apparent trend in the error. The CA method approaches the full BD simulation upon another iteration $w=2$, indicating the errors in the $w=1$ iteration arise from conformationally averaging the Brownian noise over the FD conformations. While variations in the TEA parameters are generally small, the transient conformational distributions in the $w=0$ FD iteration show significant qualitative differences relative to the $w=1$ case including HI (Supplementing Information), explaining the improvement upon a third CA iteration.

The transient extensional viscosity, shown in Figure~\ref{fig:verification_trans}b, follows the trends of the fractional extension. While there is noise in the measurement, similar fluctuations are observed even in the dilute case and there is no observable trend in the error of the CA method. Solution stress is highly sensitive to preaveraging approximations, \cite{ottinger1996stochastic, prakash1999kinetic, young2018conformationally} so this is a promising result validating the accuracy of our CA approach.

\subsubsection{Steady state and transient conformational distributions}

\begin{figure}[htb]
	\includegraphics[width=\columnwidth]{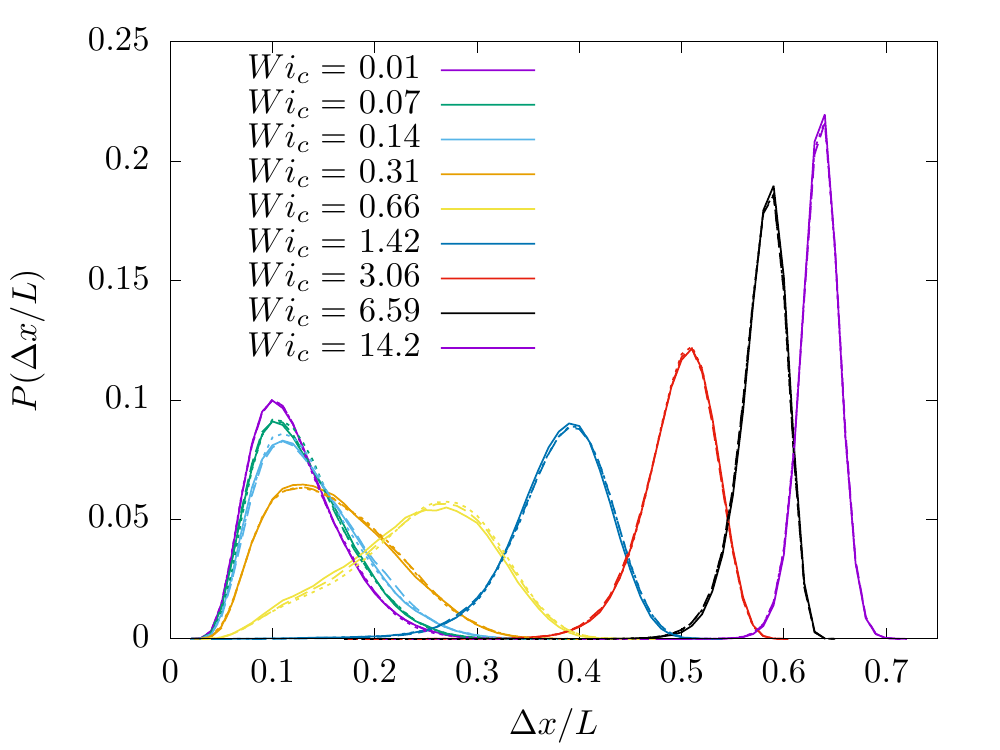}
    \caption{Steady state probability distribution functions of fractional extension for the TEA method (solid lines), the CA method $w=1$ (dashed lines), and the CA method $w=2$ (dotted lines). }
    \label{fig:verification_sspdf}
\end{figure}

While ensemble and solution average properties are useful for evaluating simulation accuracy, non-equilibrium polymer dynamics exhibit a broad range of unique conformations, often referred to as molecular individualism.\cite{de1997molecular,hur2001dynamics} This diversity can significantly modify the solution dynamics and rheology, so it is essential that the CA method also capture these effects. We calculate the steady state and transient probability distribution functions (PDF) of fractional extension $P(\Delta x/L)$ to test for these features. In the steady state case (Fig \ref{fig:verification_sspdf}), we find that the distributions for both $w=1$ and $w=2$ quantitatively match the full BD simulation similar to the ensemble average properties. 

\begin{figure}[htb]
	\includegraphics[width=\columnwidth]{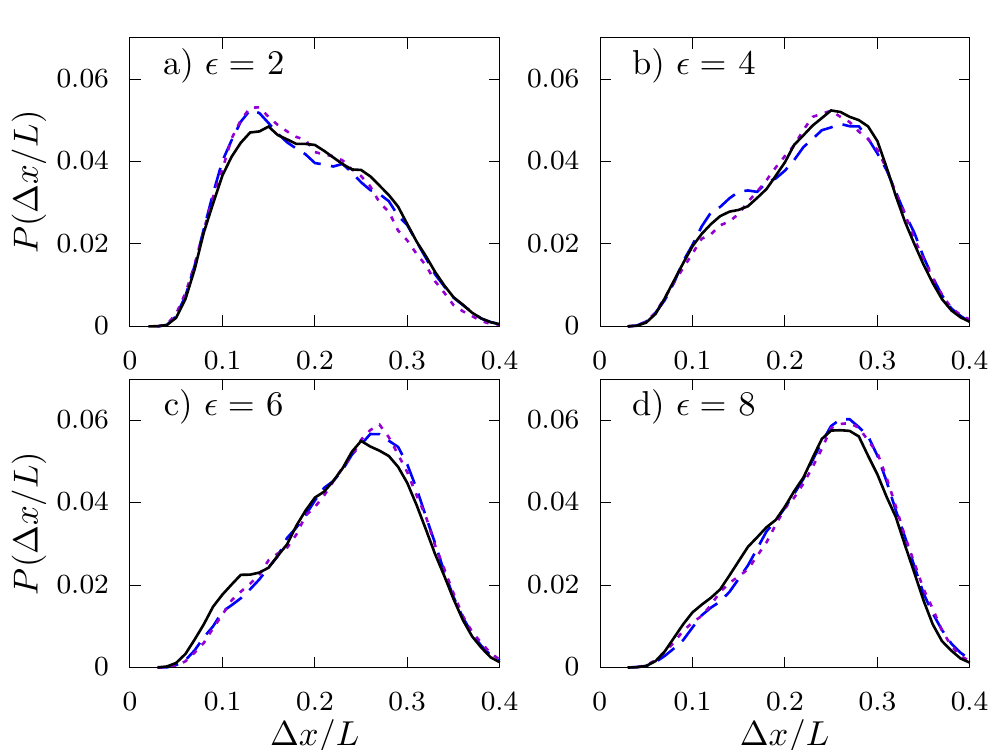}
    \caption{Transient state probability distribution functions of fractional extension with strain rate $Wi_c = 0.66$ and accumulated strain a) $\epsilon = 2$ b) $\epsilon = 4$ c) $\epsilon = 6$ d) $\epsilon=8$ for the TEA method (solid black lines), the CA method $w=1$ (dashed blue lines), and the CA method $w=2$ (dotted purple lines). }
    \label{fig:verification_transpdf}
\end{figure}

We plot the transient distributions in Figure~\ref{fig:verification_transpdf}, plotting at a strain rate and accumulated strain values where extensional fluctuations are the largest ($Wi_c=0.66$ and $\epsilon=2-8$); this should represent the least accurate situation for the CA method. Here, the lag in transient fractional extension relative to the full BD simulation is connected to individual conformations. In particular, the emergence of a steady state peak is slightly slower than in the full BD simulation as polymers remain coiled, as evidenced in Figure~\ref{fig:verification_transpdf}a and b, at $\epsilon = 2$ and $4$. However, as the chains approach steady state, the distribution matches the full BD simulation nearly quantitatively ($\epsilon=6$, Fig \ref{fig:verification_transpdf}c). Upon a third iteration $w=2$, the CA method improves further due to the refined TEA parameters. At all other flow rates the transient distributions are in near quantitative agreement for all values of accumulated strain. Because the errors in the second iteration $w=1$ are largely quantitative and do no disturb the emergence of a broad and diverse range of conformational distributions, we choose to perform only two iterations of the CA method in the following results.

\subsection{\label{sec:concdep}Concentration dependent dynamics and rheology}

We demonstrate the utility of the CA method by simulating large systems challenging to access with traditional BD simulations. We simulate chains of length $N_{b}=100$ at concentrations $c/c^{*}=0.4-3.0$ under strain rates passing through the coil-stretch transition $Wi \approx 0.1-5$. We also perform single chain dilute BD simulations without conformational averaging using the TEA for comparison. Each ensemble contains 500-800 molecules, taken from multiple simulations of $25-120$ molecules each. A representative set of simulation details at the highest and lowest strain rates are shown in Table~\ref{table:simulationparameters}.  At higher concentrations and strain rates we increase the number of chains $N_{c}$ simulated in order to maintain $l_{x} > \langle \Delta x / L \rangle_{ss}$, where $l_{x}$ is the shortest length of the simulation cell in the extension direction over the KRBC period. In principle, the simulation cell size should be at least twice the contour length, $l > 2L$, to prevent chains from interacting with themselves through the periodic boundary. In practice, this often requires a prohibitively large number of beads $N$, and previous studies have shown that the cell size has only small quantitative effects on polymer dynamics.\cite{stoltz2006concentration}

\begin{figure*}[htb]
	\includegraphics[width=\textwidth]{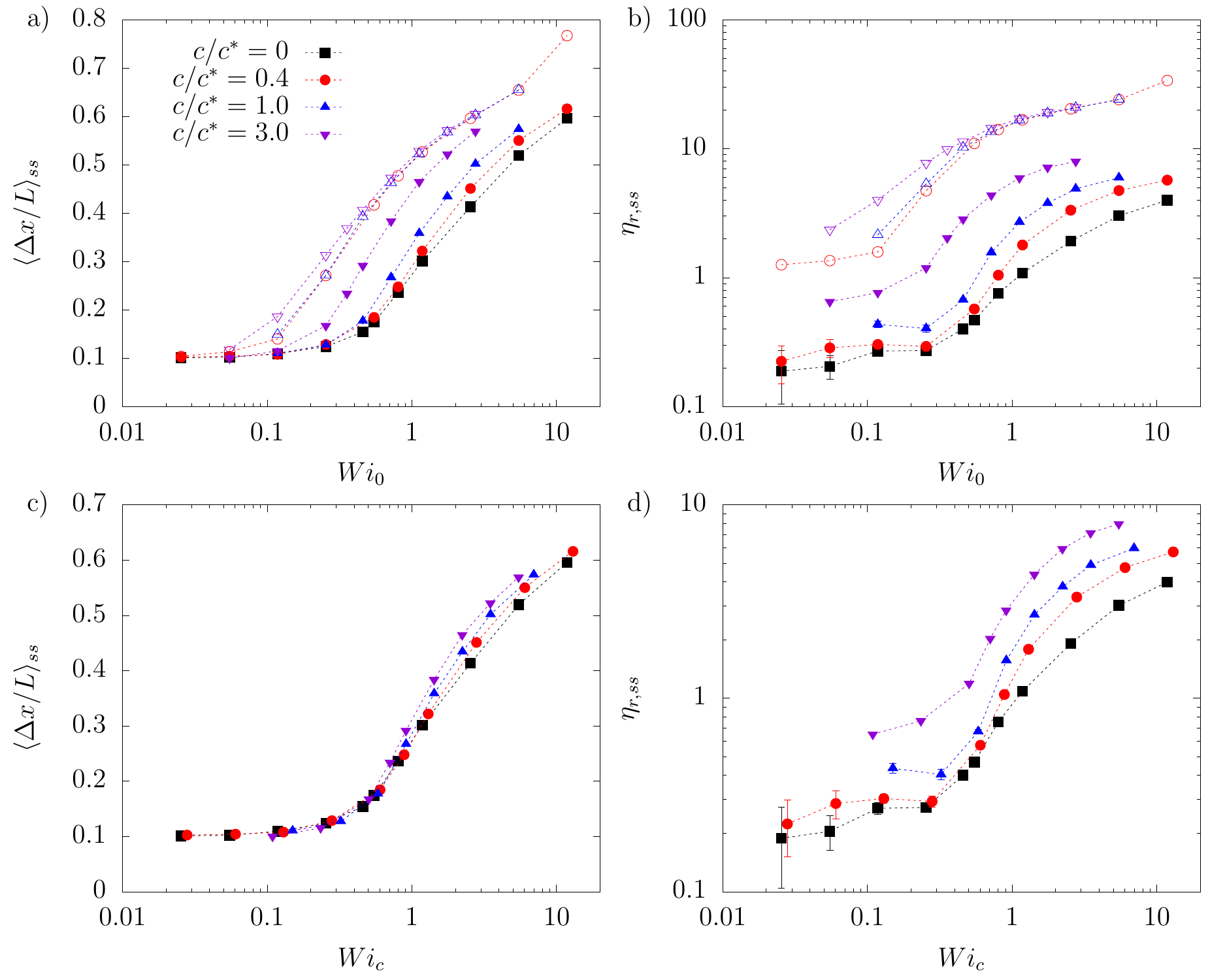}
    \caption{Steady state a) fractional extension b) extensional viscosity as a function of flow strength $Wi_0$ for various concentrations. Open symbols correspond to the first iteration of the CA method $w=0$ without HI, closed symbols to the second iteration of the CA method $w=1$ with HI, and closed black symbols to single molecule traditional BD simulations with HI. c) and d) correspond to the same data, except plotted against he concentration dependent effective flow strength $Wi_c$.}
    \label{fig:concdep_ss_Wi0}
\end{figure*}

\begin{table*}[t]
	\centering
    \begin{tabular*}{\textwidth}{c @{\extracolsep{\fill}} cccccc}
    \hline\hline
    	$c/c^*$ & $Wi_c$ & $N_c$ & $\langle \Delta x \rangle$ & $\tilde{L}$ & $\tilde{\tau}_c$\\
        \hline
        0.4 & 0.028-13.0 & 50 & 15.3-91.5 & 123.7 & 574.7\\
        1.0 & 0.15-6.95 & 25-80 & 16.4-85.3 & 70.9$-$104.5 & 645.8\\
        3.0 & 0.11-5.46 & 30-120 & 14.9-84.5 & 52.2$-$82.9 & 1008.4\\
    \hline\hline
    \end{tabular*}
    \caption{Simulation parameters for various concentrations for chains of length $N_b=100$. Ranges are given for the lowest and highest simulated flow strength $Wi_c$, and the corresponding ranges for average polymer extension in the flow direction $\Delta x$, box length at the beginning of a KRBC period $\tilde{L}$, and longest polymer relaxation time $\tilde{\tau}_c$. The dilute relaxation time is $\tilde{\tau}_Z = 508.5$.}
	\label{table:simulationparameters}
\end{table*}

\subsubsection{\label{sec:ssens}Steady state ensemble averages}

In Fig \ref{fig:concdep_ss_Wi0}a we plot the steady state fractional extension as a function of the strain rate normalized by the dilute longest polymer relaxation time from single molecule TEA BD simulations with HI, $Wi_0 = \dot{\epsilon} \tau_{Z}$. The first iteration $w=0$ of the CA method again shows that without HI polymers stretch at a low strain rate $Wi_0 \approx 0.1$. This occurs due to lack of hydrodynamic shielding, with the full Stokes drag on each monomer contributing to a long polymer relaxation time. For $w=0$, there is little concentration dependence except near the critical coil-stretch transition rate ($Wi_0 \approx 0.1-0.5$), where the fractional extension increases due to slightly higher polymer relaxation time associated with greater EV repulsions in the semidilute case. This effect is most significant near $Wi_0 \approx 0.1-0.3$ because the polymer conformation changes suddenly over a small range of strain rate. In the low $Wi$ limit, the conformation is insensitive to flow and changes in extension arise only from equilibrium chain repulsion. In the high $Wi$ limit, polymers at different concentrations approach the same extension as the finitely extensible springs require large increases in strain rate to further stretch the chain.

On the second iteration $w=1$ of the CA method, HI is included and concentration dependence increases dramatically. Even below the overlap concentration, $c/c^{*}=0.4$, the fractional extension is measurably greater than the dilute case. This is consistent with experimental and simulations observations that polymer solution dynamics and rheology becomes concentration dependent significantly below the equilibrium overlap concentration when an extensional flow is imposed.\cite{clasen2006dilute} At the overlap concentration $c/c^{*}=1.0$ the fractional extension increases by as much as a factor of 1.3 at the critical coil-stretch transition rate, $Wi_0 = 0.5$. Above the overlap concentration, $c/c^{*}=3.0$, the increase is as large as a factor of 2 at $Wi_0 = 0.5$. Additionally, higher concentration solutions stretch at a lower strain rate, and the $w=1$ simulation results approach the $w=0$ results. This can again be explained by an increase in polymer relaxation time, except now the effect is much greater because in the dilute case HI strongly shields the polymer from solvent drag. At higher concentrations, HI becomes screened and the polymer is exposed to a higher effective flow strength for the same applied strain rate.

Next we consider the steady state extensional viscosity (Fig \ref{fig:concdep_ss_Wi0}b). There is a greater concentration dependence with similar qualitative trends as the fractional extension results. The viscosity measured from FD simulations is significantly greater than all HI simulations, even upon accounting for a linear concentration dependence. This is consistent with the previous description of enhanced flow penetration and thus greater polymer contribution to the stress (Fig \ref{fig:verification_ss}b), even in the equilibrium case where the polymer remains coiled in both freely draining and hydrodynamically interacting simulations. 

When HI is included, the strain rate range of concentration dependent rheology extends below the coil-stretch transition. For $c/c^{*}=3.0$, HI screening causes flow to penetrate the coil as low as $Wi_0 = 0.13$, leading to an increase in polymer stress relative to the dilute case, even at similar fractional extension. The change is greatest near the dilute coil-stretch transition, $Wi_0=0.5$, where the viscosity increases by a factor of $7.5$ in the $3c^{*}$ solution. At higher flow rates, viscosity plateaus in the semidilute solution as the polymers become completely stretched and exposed to flow. The concentration dependence is maintained to high strain rates where the polymers are fully aligned in the flow direction, however.

\begin{figure*}[htb]
	\includegraphics[width=\textwidth]{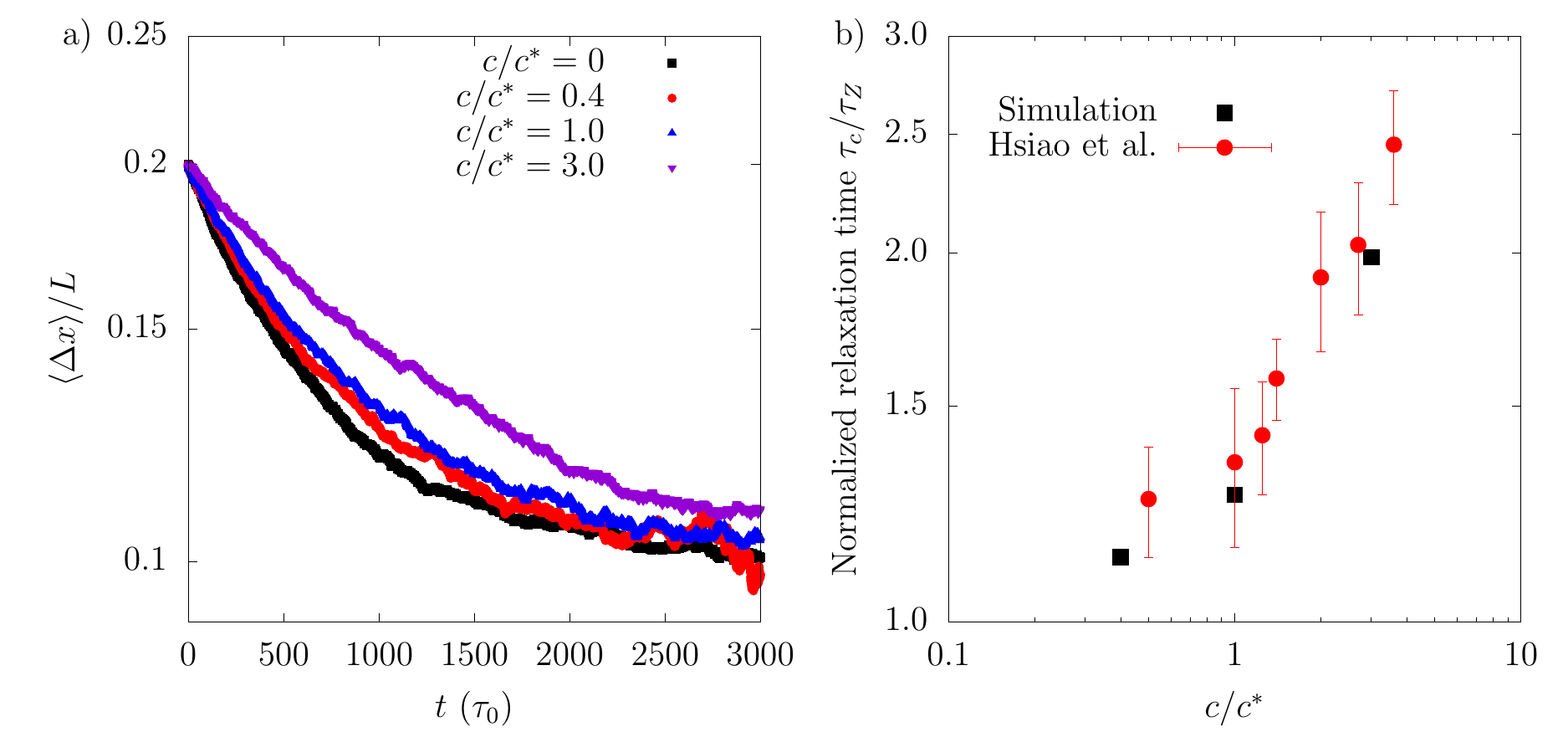}
    \caption{a) Average fractional extension $\langle \Delta x \rangle/L$ as a function of time $t$ and polymer concentration. Polymer chains are initially stretched via planar extensional flow, which is turned off at $t=t_{cess}$. The relaxation of $\langle \Delta x \rangle/L$ occurs over a characteristic time $\tau_c$, which we obtain by fitting the data after time $t=0$ where $\langle \Delta x \rangle / L = 0.2$. We then normalize by the dilute limit relaxation time $\tau_z$ and plot as a function of concentration $c/c^*$ in b), along with experimental data from Hsiao, et al.\cite{hsiao2017direct} Values of $\tau_c/\tau_Z$ are used to normalize the Weissenberg number $Wi_c$ and exhibit quantitative agreement with experiment within error. Each concentration represents an emsemble average of 500 molecules.}
    \label{fig:ctau}
\end{figure*}

Both the FD and HI results are consistent with previous findings of Stoltz et al. \cite{stoltz2006concentration} from coarse-grained bead-spring simulations of semidilute solutions of DNA. Following their approach, we now rescale all strain rates by longest polymer relaxation time at the concentration of interest, $Wi_c = \dot{\epsilon}\tau_c$ (Fig.~\ref{fig:concdep_ss_Wi0}c,d). The dynamics and rheology are then compared on the basis of the effective flow strength and not the solvent deformation rate $\dot{\epsilon}$ as in Fig \ref{fig:concdep_ss_Wi0}a,b. Relaxation times were obtained as in Section \ref{sec:Verification}. Relaxation curves at $c/c^*=0,0.4,1.0,3.0$ are shown in \ref{fig:ctau}a, and corresponding relaxation times normalized by the Zimm relaxation time $\tau_Z$ determined from dilute simulations are plotted versus concentration in \ref{fig:ctau}b. We have compared to the single molecule experiments of Hsiao et al. \cite{hsiao2017direct} on $\lambda$-DNA and found good agreement. Below the overlap concentration, $\lambda$-DNA shows a larger relaxation time than our simulations. This is expected, as we have simulated a shorter polymer ($N_b=100$ at 1.8 beads per persistence length, versus 200 persistence length $\lambda$-DNA). Longer polymers experience self-concentration effects that lead to longer relaxation times at lower normalized concentrations $c/c^*$ as compared to shorter polymers. \cite{clasen2006dilute, prabhakar2016influence} However, these differences are smaller at higher concentrations, and in all cases the simulation results and within the error bars of the experimental values. The simulations contain ensembles of 500 molecules, so we find the error bars are smaller than the point size.

When comparing results on the basis of $Wi_c$, we omit the freely draining results from the $w=0$ iteration of the CA method for clarity. The FD results can be collapsed onto the HI simulations in the low and high $Wi_c$ limits using the appropriate relaxation time measured from FD simulations, however only the simulations including HI capture experimentally relevant behavior. Upon rescaling, the fractional extension concentration dependence is almost completely eliminated. For flow strengths up to $Wi_0 \approx 1.0$, the longer relaxation time at higher concentration cancels out the increased polymer drag. At higher flow strengths, extension increases slightly with concentration. We suggest this occurs because the increase in drag with concentration is greater for a stretched polymer than for a coiled polymer;\cite{prabhakar2016influence} however, these differences are sufficiently small that simulations of longer polymers at higher concentrations may be necessary to fully characterize this effect.



To contrast, the extensional viscosity shows a significant concentration dependence upon rescaling to $Wi_c$. Viscosity increases in all semidilute solutions starting at the critical coil-stretch transition rate and continuing to high flow strengths. Even at a lower applied strain rate, the lowest concentration $c/c^{*}=0.4$ shows a factor of 2 increase in viscosity due to enhanced solvent drag caused by HI screening. The location of the coil-stretch transition does not change when increasing concentration to $c/c^{*}=3.0$, but the viscosity is significantly greater than the dilute case for all flow strengths.


\subsubsection{\label{sec:trens}Transient ensemble averages}

\begin{figure*}[htb]
	\includegraphics[width=\textwidth]{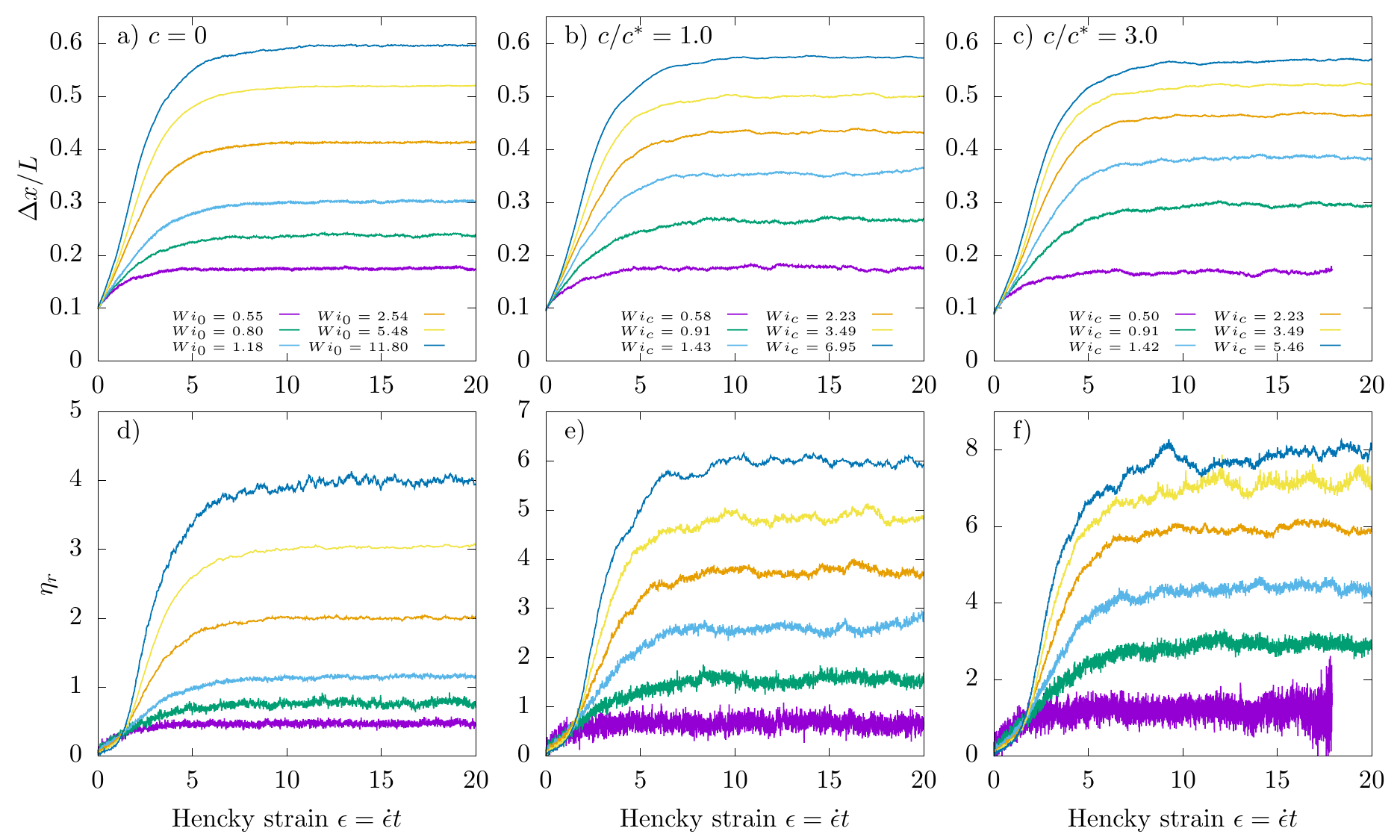}
    \caption{Top panels show the transient fractional extension for $N_{b}=100$ and varying flow rate $Wi$ at a concentration a) $c/c^{*} = 0$ b) $c/c^{*} = 1.0$ c) $c/c^{*} = 3.0$. Bottom panels show the transient extensional viscosity at the same conditions as the panel above.}
    \label{fig:transient_ensavg}
\end{figure*}

The transient response of polymers in the startup of extensional flows is also of interest. Hsiao et al. and Samsal et al. found that the average transient extension in dilute and semidilute solutions at the same $Wi_c$ are nearly identical.\cite{sasmal2017parameter,hsiao2017direct} We plot the same quantities in Figure~\ref{fig:transient_ensavg}a-c, which show representative transient fractional extension $\Delta x/L$ as a function of Hencky strain $\epsilon = \dot{\epsilon} t$. Our findings are similar to Hsiao, et al. and Sasmal, et al.\cite{hsiao2017direct,sasmal2017parameter} in that there are generally no qualitative differences between dilute and semidilute transient fractional extension.

Despite the similarities between dilute and semidilute solutions in the transient fractional extension $\Delta x/L$, marked differences are observed in the transient extensional viscosity $\eta_r$ for the same systems. $\eta_r$ is plotted in Figure~\ref{fig:transient_ensavg}d-f as a function of $\epsilon$. Consistent with the prior results in Fig.~\ref{fig:concdep_ss_Wi0}, increased $Wi_c$ results in an increasingly large steady-state value of $\eta_r$. Interestingly, given a comparable number of molecules in the ensemble average (ca 500-800 for each concentration and strain rate), semidilute solutions exhibit larger fluctuations in extensional viscosity (Fig.~\ref{fig:transient_ensavg}d-f) with increasing concentration. This includes significant fluctuations at the highest flow rates ($c.c^{*}=3.0, Wi_c=3.49-5.46$ Fig \ref{fig:transient_ensavg}f) that occur over multiple strain units $\epsilon$ due to the rapid pulling rate (i.e. high $\dot{\epsilon}$). In part, we attribute the emergence of these large fluctuations due to the sensitivity of the force-extension relation at high strech, where small fluctuations in bond stretch can lead to large changes in the force and thus viscosity. This is supported by the observation that fluctuations in fractional extension at these conditions are not measurably larger than lower strain rates.

While there are indeed small fluctuations in the average fractional extension $\Delta x/L$ versus $\epsilon$ during transient stretching, the highly-fluctuating extensional viscosity $\eta_r$ reveals the importance of individual polymer dynamics. Indeed, while our ensemble average results are consistent with previous simulation and experimental studies, we can show that conformational fluctuations, molecular trajectories, and topological interactions play a key role in the measurable dynamics of semidilute solutions.

\begin{figure*}[htb]
	\includegraphics[width=\textwidth]{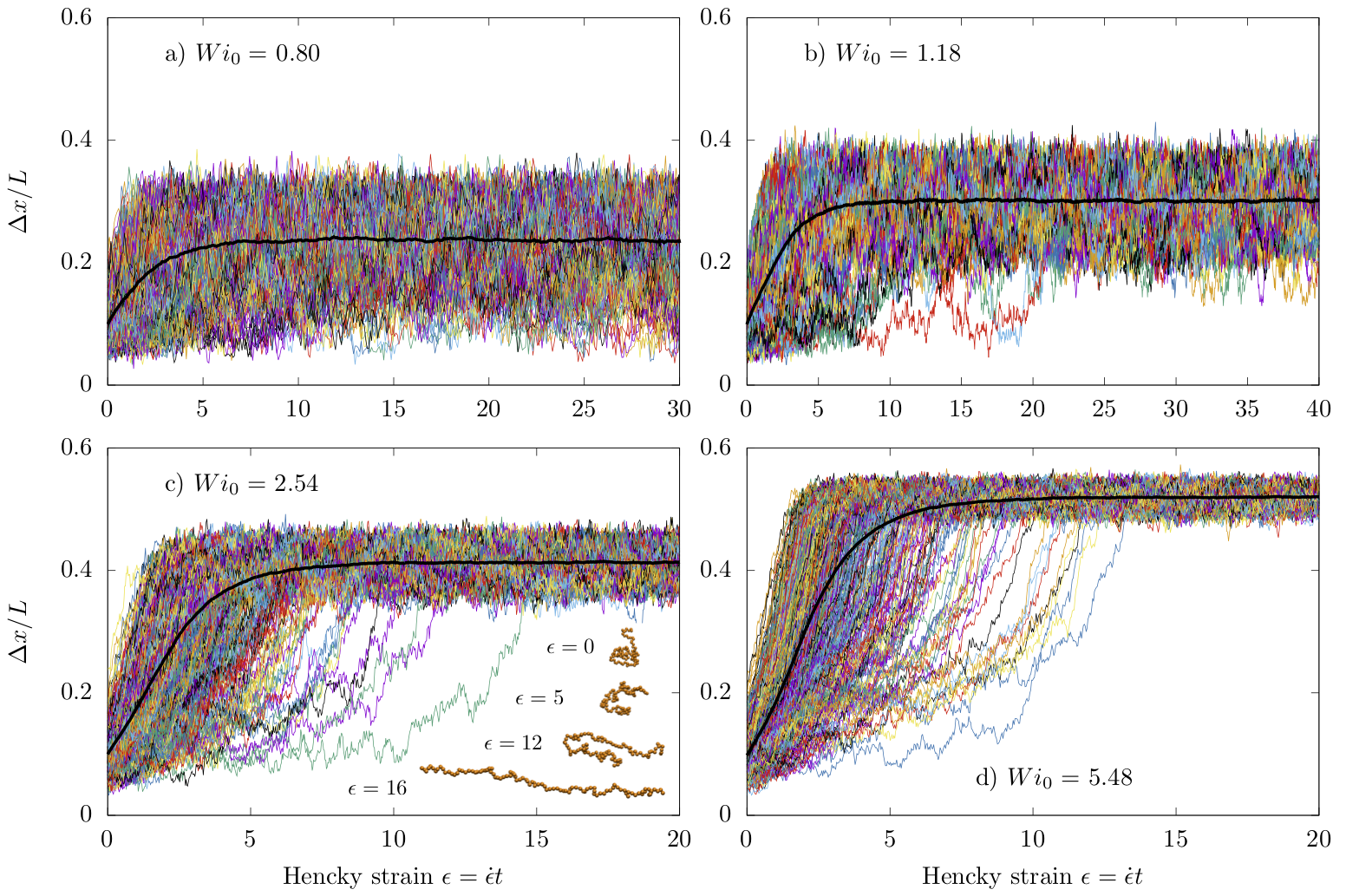}
    \caption{Individual molecular trajectories (thin colored lines) and ensemble averages (thick black lines) of transient fractional extension for dilute polymer solutions at increasing flow strengths a) $Wi_0=0.80$ b) $Wi_0=1.18$ c) $Wi_0=2.54$ d) $Wi_0=5.48$. Inset of c) shows an example of a slow stretching conformation at $Wi_0 = 2.54$ for the green trajectory.}
    \label{fig:transient_traces0}
\end{figure*}

\subsubsection{\label{sec:trajectories}Molecular trajectories}

With the computational speedup of the CA method, we can simulate polymers at the relatively fine grained level required to probe topological interactions and conformational distributions while still accessing experimentally relevant time and length scales. For example, the experimental results of Hsiao et al. \cite{hsiao2017direct} reveal broad transient distributions and evidence of flow-induced entanglements in the startup of planar extension. The scaling theory of Prabhakar et al. \cite{prabhakar2016influence} for unentangled semidilute solutions in extensional flows predicts concentration-dependent conformational distributions and hysteresis. Broad conformational distributions and hysteresis are typically associated with the conformation dependent HI of a polymer, \cite{de1974coil, schroeder2003observation} raising questions as to the influence of concentration dependent HI screening on this phenomena. Using the CA method, we can perform detailed investigations of molecular quantities generally inaccessible via simulation. In the remainder of this article, we make several such observations.

We plot individual molecular trajectories of transient fractional extension along with the ensemble averages in Fig \ref{fig:transient_ensavg}a-c. At each concentration $c/c^{*}=0,1.0,3.0$ we select four flow strengths of interest beginning near the critical coil-stretch point and increasing to flow strengths where all polymers are aligned in the flow direction at steady state.

\begin{figure*}[htb]
	\includegraphics[width=\textwidth]{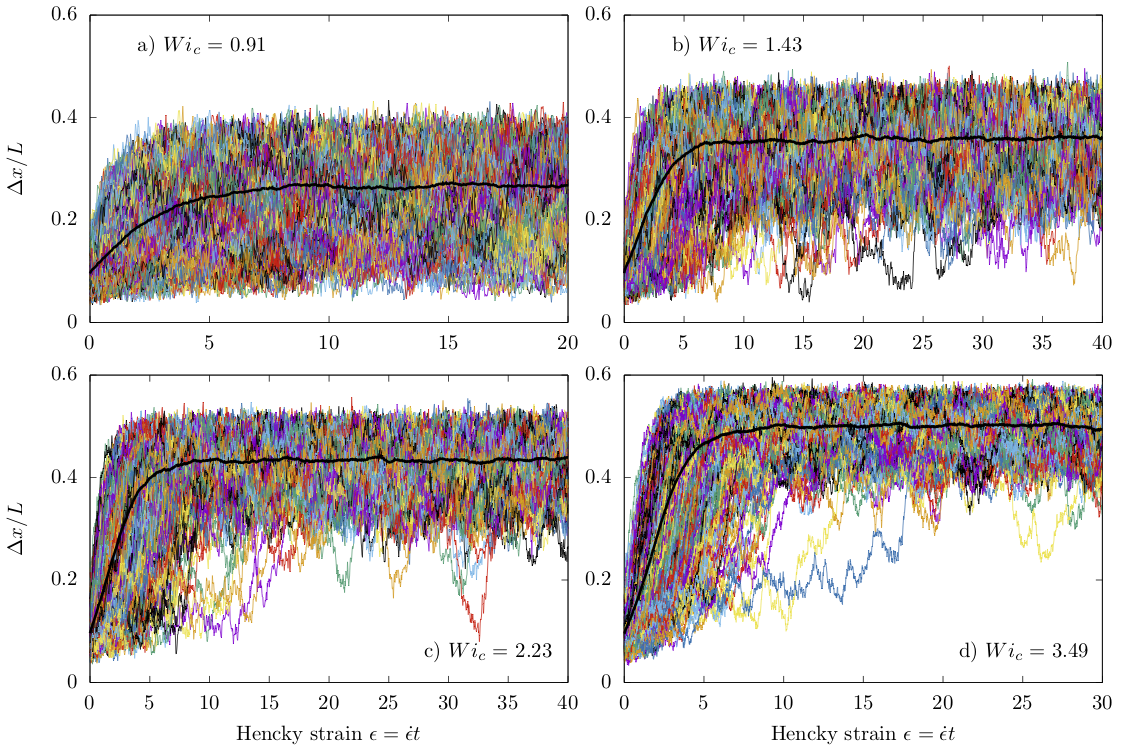}
    \caption{Individual molecular trajectories (thin colored lines) and ensemble averages (thick black lines) of transient fractional extension for semidilute polymer solutions at a concentration $c/c^{*}=1.0$ and increasing flow strengths a) $Wi_c=0.91$ b) $Wi_c=1.43$ c) $Wi_c=2.23$ d) $Wi_c=3.49$.}
    \label{fig:transient_traces1}
\end{figure*}

The startup response of dilute polymer solutions to planar extensional flow has been well studied by experiment, simulation, and theory.\cite{babcock2003visualization,perkins1997single,schroeder2003observation} In Fig \ref{fig:transient_traces0} we present simulation traces of extension $\Delta x/L$ versus Hencky strain $\epsilon$, which are similar to previous work,\cite{perkins1997single} to directly compare with the semidilute case for an identical polymer model. At relatively low flow strengths just past the critical coil-stretch rate, $Wi_0=0.80$, polymers are disturbed from their equilibrium conformations and stretch in the flow direction. The flow is not strong enough to sustain stretched conformations, leading to large conformational fluctuations in the range of $\Delta x/L = 0.03-0.35$ characterized by a broad distribution of extensions. Under stronger flows, $Wi_0 = 1.18$, polymers stretch to a higher extension and experience smaller fluctuations around the average. There are infrequent fluctuations towards lower extension, but polymers do not return to equilibrium coiled levels of extension. 

Further increasing flow strength, $Wi_c=2.54-5.48$, leads to higher extensions and smaller fluctuations around the steady state average. We note molecular individualism in a small population of trajectories which stretch much slower than the ensemble average and do not reach steady state until $\epsilon=10-15$. These folded conformations (an example of which is shown in the inset of Fig. \ref{fig:transient_traces0}c) likely form due to the local stability of the hairpin conformation, which which has both chain ends being stretched in the same direction. 

At the overlap concentration (Fig \ref{fig:transient_traces1}), molecular trajectories at low flow strengths, $Wi_c=0.91$, are visually similar to the dilute case. Polymers undergo large conformational fluctuations around the steady-state fractional extension as the flow is strong enough to deform the equilibrium structure, but not sufficient to sustain stretched conformations. At high flow strengths, $Wi_c=3.49$, trajectories are also qualitatively similar to the dilute case in that polymers stretch to a steady state conformation and do not return to the coiled state, with some slowly-stretching folding conformations. Nevertheless, we do note some quantitative differences evident in the large fluctuations around the steady state extension relative to the dilute case.

\begin{figure*}[htb]
	\includegraphics[width=\textwidth]{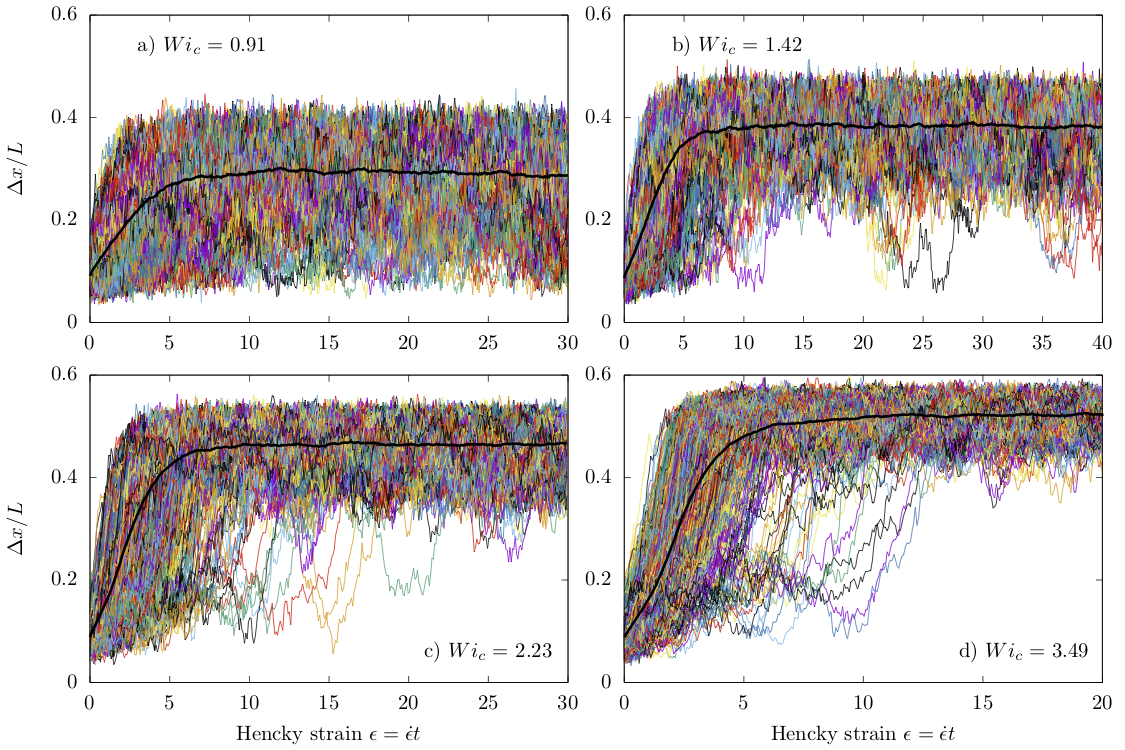}
    \caption{Individual molecular trajectories (thin colored lines) and ensemble averages (thick black lines) of transient fractional extension for semidilute polymer solutions at a concentration $c/c^{*}=3.0$ and increasing flow strengths a) $Wi_c=0.91$ b) $Wi_c=1.42$ c) $Wi_c=2.23$ d) $Wi_c=3.49$.}
    \label{fig:transient_traces3}
\end{figure*}

At intermediate flow strengths, $Wi_c=1.42-2.23$, fluctuations are larger, leading to broad distributions of fractional extension relative to the dilute case. In addition to this quantitative change, semidilute trajectories display qualitatively different features. After reaching a moderately stretched and broadly distributed steady state value, a small population of polymers returns to the coiled conformation. This is contrary to the dilute case, where polymers reaching comparable steady state fractional extensions do not return to the coiled state. Generally these coiled conformations are not long-lived, and the polymers restretch. This suggests the dynamic free energy barrier going from the coiled to stretched conformation is small,\cite{de1976dynamics} leading to interconversion between coiled and stretched states but not a coexistence of two stable populations. We include a simulation movie demonstrating an example of coil-stretch interconversion for $c/c^{*}=1.0, Wi_c=1.43$ in the Supplementary Information.

Above the overlap concentration at $c/c^{*}=3.0$ (Fig \ref{fig:transient_traces3}), molecular trajectories are qualitatively similar to the overlap case. Low and high flow strength stretching behavior is again similar to dilute solutions, with large conformational fluctuations near the critical coil-stretch rate $Wi_c=0.91$ and uniform stretching in the flow direction for strong flows $Wi_c=3.49$. At intermediate flow strengths, a window emerges where polymers undergo interconversion between coiled and stretched states. We observe that coiled conformations are less common and shorter-lived than at the overlap concentration (Fig. \ref{fig:transient_traces1}). 

\subsubsection{\label{sec:sspdf}Steady state conformational distributions}

While molecular trajectories are useful for observing qualitative differences upon increasing concentration, they do not quantify the change in polymer conformations. To achieve this, we consider steady state and transient probability distribution functions (PDFs) of polymer fractional extension. As demonstrated in our verification of the CA method in Section \ref{sec:Verification}, distributions of extension provide a detailed and quantitative analysis of polymer dynamics. Additionally, these quantities are directly comparable to results from single molecule experiments. 

\begin{figure*}[htb]
	\includegraphics[width=\textwidth]{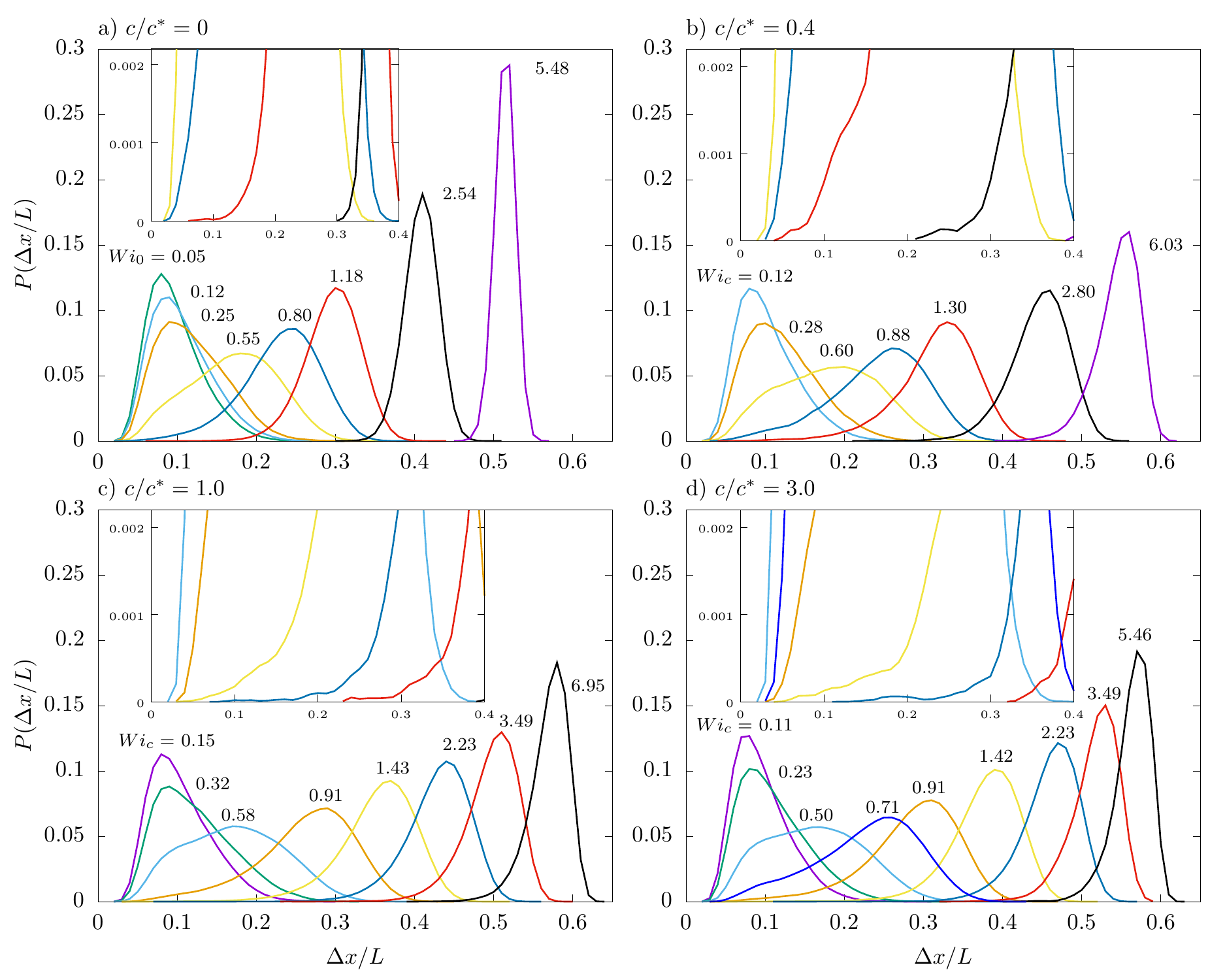}
    \caption{Steady state probability distribution functions of fractional extension at concentrations a) $c/c^{*}=0$ b) $c/c^{*}=0.4$ c) $c/c^{*}=1.0$ d) $c/c^{*}=3.0$ as a function of flow strength. Insets are zoomed in views of selected distributions}
    \label{fig:sscpdf}
\end{figure*}

In Fig \ref{fig:sscpdf}, we plot steady state distributions for the molecular trajectories shown in Fig \ref{fig:transient_traces0}-\ref{fig:transient_traces3}. We define steady state to occur when all polymers in a given ensemble have stretched to the high extension peak in the PDF. This includes polymers which retract to coiled conformations at high flow rates when $\epsilon>\epsilon_{ss}$, although for $Wi_c>0.5$, we do not observe any polymers which remain coiled for the entire simulation. Generally, we find this steady state to occur at $\epsilon_{ss}=15$. 



For dilute solutions, the shape of the distribution changes sharply near the critical coil-stretch rate. At low flow strengths, polymers are narrowly distributed around the equilibrium stretch with relatively little rate dependence from $Wi_0=0.05-0.25$. At the coil-stretch transition, $Wi_0=0.55$, the distribution is the broadest corresponding to large fluctuations between equilibrium coiled and non-equilibrium stretched conformations. At higher flow strengths, distributions shift to higher extension and rapidly narrow. 

Semidilute conformational distributions contrast strongly, quantifying observations from molecular trajectories in Figures~\ref{fig:transient_traces0} to~\ref{fig:transient_traces3}. At the overlap concentration $c/c^{*}=1.0$, distributions at all flow strengths above $Wi_c=0.5$ are significantly broader. This behavior is particularly evident at high flow strengths, where distributions narrow only slightly as the peak in extension shifts to the right. We also include insets that zoom in on the region where coil-stretch interconversion is expected, $\Delta x/L = 0-0.4$. While fewer than 1\% of polymers are coiled, $P(\Delta x/L) \approx 5\times 10^{-3}$, the qualitative differences relative to the dilute case are apparent. At the overlap concentration, low and equilibrium-like values of $\Delta x/L$ are observable for flow strengths as high as $Wi_c=1.36$. In the dilute case, polymers do not fluctuate back to equilibrium levels of extension at a comparable flow strength $Wi_0 = 1.18$. 

The shape of the distributions is also of note. Broad distributions are not well described by a Gaussian, the prediction of kinetic theory for dilute polymer solutions in extensional flows.\cite{bird1987dynamics2} The semidilute distributions are highly asymmetric, with the low-extension tail extending further from the peak than the high-extension tail. This trend continues until strong flows, $Wi_c>3$, where polymers become completely aligned in the flow direction. In comparison, the dilute solution distributions are nearly Gaussian for all flow strengths except the critical coil-stretch transition rate, where kinetic theory preaveraging approximations break down even in dilute solutions. 

As in our discussion of molecular trajectories, distributions above the overlap concentration $c/c^{*}=3.0$ are qualitatively similar to $c/c^{*}=1.0$. It is notable, however, that distributions at the same effective flow strength $Wi_c$ are noticeably narrower. The steady state coiled population at intermediate flow strengths also appears to be reduced, but quantitative conclusions cannot be made here as the number of coiled polymers at steady state is small relative to the size of the ensemble.

The qualitative difference in semidilute distributions cannot be explained by the higher solvent deformation rate $\dot{\epsilon}$ at the same effective flow strength $Wi_c$ alone. Hydrodynamic interactions between polymers at varying levels of stretch are of central importance here. We speculate that the broadened distributions emerge due to the conformation dependent drag on a polymer, particularly the difference between drag on a stretched and coiled polymer in semidilute solution. 

\begin{figure*}[htb]
	\includegraphics[width=\textwidth]{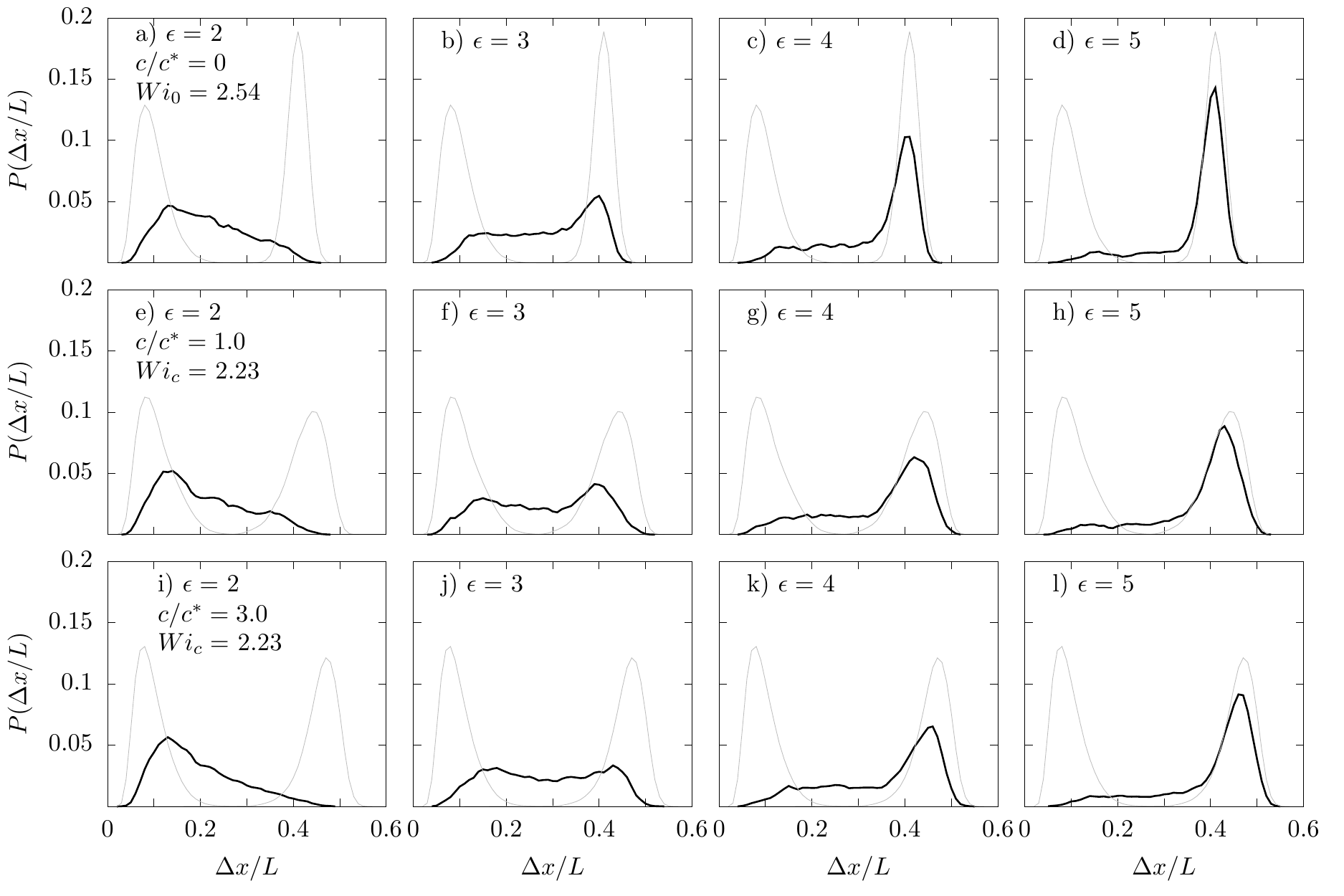}
    \caption{Transient probability distribution functions of fractional extension at a-d) $c/c^{*}=0, Wi_0 = 2.54$ e-h) $c/c^{*}=1.0, Wi_c=2.23$ i-l) $c/c^{*}=3.0, Wi_c=2.23$ as a function of increasing accumulated strain $\epsilon = 2-5$. Grey lines at low and high extension represent the equilibrium and steady state distributions respectively.}
    \label{fig:ssctpdf}
\end{figure*}

Our results are qualitatively comparable to the theoretical predictions of Prabhakar et al. for semidilute unentangled melts.\cite{prabhakar2016influence} In this study, it was found that the self-concentration effect below $c^{*}$ increases drag on stretched polymers, while coiled polymers remain nominally shielded from flow because the HI correlation length is similar to the size of the polymer.\cite{prabhakar2016influence} While we do not see a hysteresis window due to our relatively short chain length, a broad conformational distribution and the interconversion between coiled and stretched states is consistent with the results of conformation dependent drag and self concentration.

Above the $c^{*}$, Prabhakar et al. predict the hysteresis window for high molecular weight polymers decreases.\cite{prabhakar2016influence} In this case, stretched polymers self-dilute the solution and reduce intermolecular interactions compared to equilibrium.\cite{prabhakar2016influence} This is again consistent with our simulation results in that we observe conformational distributions above the overlap concentration become sharper and more Gaussian, with a decrease in the coiled population. 

\subsubsection{\label{sec:trpdf}Transient conformational distributions}

The molecular trajectories also suggest that the transient behavior of semidilute solutions is more diverse than in the dilute case, particularly for strong flows where polymers are quickly deformed from their equilibrium conformations. Importantly, we also expect that the effect of topological interactions will be greatest in the startup of flow. To quantify this observation, we plot transient distributions in Figure~\ref{fig:ssctpdf} as a function of accumulated strain $\epsilon=2-4$ for $c/c^{*}=0,1.0,3.0$ at comparable effective flow strengths $Wi_c \approx 2.2$. 

At all concentrations, the distributions shortly after the startup of planar extensional flow, $\epsilon = 2$, are similar. This is expected as the polymers have not yet been significantly deformed from equilibrium. By $\epsilon=3$, however, we observe subtle differences. The steady state peak has emerged in dilute solutions, and the distribution of unstretched molecules below $\Delta x / L \approx 0.3$ is nearly uniform. In semidilute solutions, a weak peak at high extension emerges, but it has not yet reached the steady value. Additionally, the unstretched population is not uniform, with another weak peak corresponding to coiled and folded molecules which remain relatively unperturbed from their equilibrium conformations. 

As strain is further accumulated, the distributions at all concentrations become qualitatively similar, with the exception of the sharper distribution in the dilute case as noted in the steady state discussion. The high extension peak in semidilute solutions continues to shift to the right and does not reach the steady state location by $\epsilon=5$, in contrast to dilute solutions where the high extension peak reaches its steady state location by $\epsilon=4$ and only sharpens thereafter.

We again conjecture that this concentration dependence is in part due to the conformation dependent drag on a polymer in semidilute solution. Upon the startup of flow, some polymers are moderately stretched due to fluctuations in conformation at equilibrium. These chains are exposed to the applied solvent deformation and solvent velocity perturbations from interchain HI, causing them to stretch faster than the ensemble average. Polymers that are coiled or folded upon the startup of flow remain shielded because the HI correlation length is comparable to the polymer size. As the solution is further deformed, HI shielding is not sufficient to prevent the coiled subpopulation from stretching, and the distribution shifts towards a single peak at high extension. However, we also note interesting conformational dynamics which are not consistent with these effectively modified equilibrium theoretical predictions, and which could explain the emergence of distinct molecular populations.

\begin{figure*}[htb]
	\includegraphics[width=\textwidth]{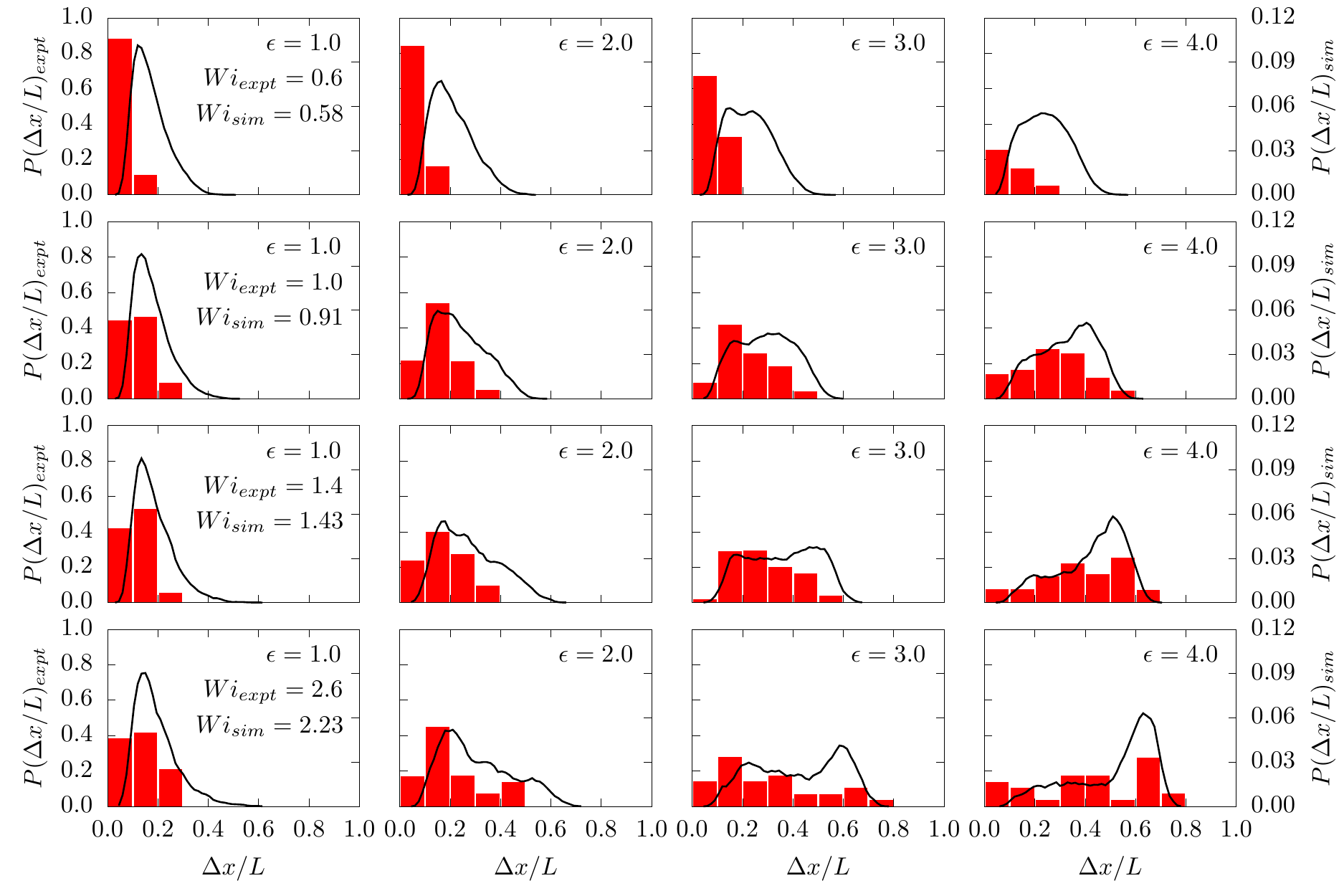}
    \caption{Transient probability distribution functions of fractional extension at the overlap concentration $c^*$ for increasing strain $\epsilon=1-4$ and flow strength $Wi_c=0.58-2.23$. We have coplotted the experimental results of Hsiao et al. \cite{hsiao2017direct} (red bars) with simulation results (black lines) for comparable flow strengths. Simulation results are plotted against a second y-axis because we use a fractional extension bin size of 0.01, whereas Hsiao et al. use 0.1. Additionally, in this figure we consider a contour length $L=(N_b-1)r_{eqm}$, where $\tilde{r}_{eqm}=1.92$ is the equilibrium average bond length. This rescales the x-axis by a factor of $r_{max}/r_{eqm}$ relative to Fig. \ref{fig:ssctpdf}.}
    \label{fig:tpdfexp}
\end{figure*}

In discussing increased molecular individualism, we again highlight the results Hsiao et al., who found that semidilute distributions of polymer extension in the startup of planar extension were broader than in dilute solutions. In particular, for a solution at the overlap concentration and effective flow strength $Wi_c=2.6$, they observed a persistence of low extension molecules up to $\epsilon=4.0$. In dilute solution at a flow strength $Wi_0 = 2.0$, all molecules had reached an extension of at least $\Delta x / L > 0.4$ by an accumulated strain of $\epsilon=4.0$. We make a direct comparison in Figure~\ref{fig:tpdfexp} with their results by coplotting transient PDFs at the overlap concentration $c^*$ for $Wi_c=0.58-2.23$ and increasing strain.\cite{hsiao2017direct} We find qualitative agreement, with both experimental and simulation results exhibiting broader distributions at $c^*$ than in the dilute case for the same flow strength $Wi_c$.

For this comparison, we consider the contour length in simulation to be $L=(N_b-1)r_{eqm}$, where $\tilde{r}_{eqm}=1.92$. The x-axis in Fig. \ref{fig:tpdfexp} is then rescaled by a factor of $r_{max}/r_{eqm}=1.56$ relative to Fig. \ref{fig:ssctpdf}. This provides a more appropriate quantitative comparison with experiment, where segments are rigid. In simulation, we observe that the stiff springs rarely stretch beyond the equilibrium length when HI is included. Using this definition, the only case where $\langle \Delta x/L \rangle >1.0$ is in the FD simulations for the highest flow strength, $Wi_c=13.0, c/c^*=0.4$. Thus the comparison is based on deformation from the equilibrium polymer conformations. A more quantitative verification requires consideration of a specific force law, which have not made in this work.

Hsiao et al. quantified the connection between distinct conformations on startup and diverse stretching behavior by dividing molecules into subpopulations. Four cases were identified: polymers that stretched uniformly, end-coiled polymers that stretched faster than the ensemble average, end-coiled polymers that stretched slower than the ensemble, and polymers that remained coiled. They suggested the end-coiled fast population could be caused by transient flow-induced entanglements of the folded portion of the chain with surrounding chains. We conjecture that the broadened conformational distributions are due in part to these topological interactions and the corresponding existence of molecular subpopulations. This is a departure from scaling and mean field theories, which do not account for the distribution of constraints from neighboring chains due to different initial conformations. Inspired by their approach, we make a preliminary investigation of molecular individualism for semidilute polymer solutions in planar extensional flow.

\subsection{\label{sec:hooking}Intermolecular hooking}

\begin{figure*}[htb]
	\includegraphics[width=\textwidth]{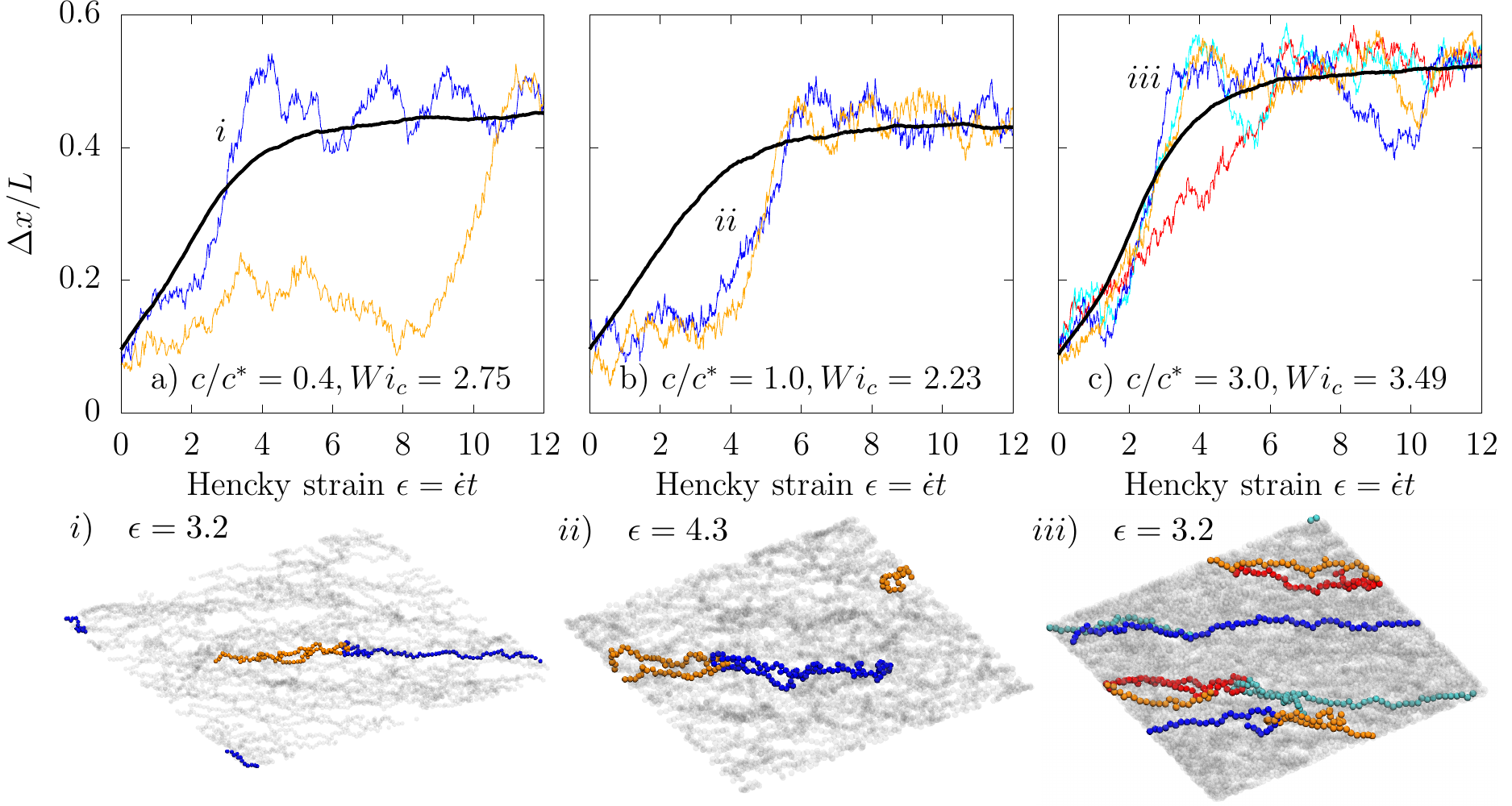}
    \caption{Simulation snapshots (bottom) and individual fractional extension trajectories of the highlighted polymers (top) demonstrating intermolecular hooking a,i) $c/c^*=0.4,Wi_c=2.75$ b,ii) $c/c^*=1.0, Wi_c=2.23$, c,iii) $c/c^*=3.0, Wi_c=4.39$. Colors of extension traces correspond to the polymer in the snapshot, and the placement of roman numerals indicates the Hencky strain at the instant of the snapshot.  }
    \label{fig:hooks!}
\end{figure*}

Here we demonstrate an essential feature of the CA method in that we are able to simulate polymers with strong excluded volume that prevent chain crossings. The polymer model simulated is commonly employed in simulations of entangled polymer melts to enforce entanglement constraints. \cite{kroger1997polymer, xu2018molecular, o2018relating} The authors have previously found that polymers of length $N_b=100$ do not exhibit entanglement dynamics at equilibrium for the highest concentration simulated here, $c/c^{*}=3.0$. \cite{young2018conformationally} Typically the entanglement concentration $c_e$ at which equilibrium polymer relaxation and diffusion dramatically slow down is in the range of $c_e = 3-5c^{*}$, although this value is not universal. \cite{rubinstein2003polymer}

Despite the fact that these solutions are nominally unentangled at equilibrium, we observe transient intermolecular hooks on startup of  planar extensional flows at concentrations as low as $c/c^{*}=0.4$ (Fig \ref{fig:hooks!}a). This phenomena is relatively rare in such dilute solutions, however, with only approximately one percent of polymers forming hooks, $P_{hook} \approx 0.01$. As concentration increases (Fig.~\ref{fig:hooks!}b and c), so does the interpenetration of equilibrium polymer coils and thus the probability of hooking. At $c/c^{*}=3.0$, the population of hooked molecules is as high as $P_{hook} \approx 0.1$. 

Additionally, hooks are empirically more common at intermediate flow strengths $\geq Wi_c \approx 1.0-3.0$. This is expected, as equilibrium entanglement dynamics emerge only when constraints from surrounding chains are dense enough that the polymer effectively reptates along its backbone contour. \cite{doi1988theory, rubinstein2003polymer} In a solution below the entanglement concentration, these obstacles are relatively dilute, and the chain can relax constraints more easily than it can reptate. When polymer conformations are deformed out of equilibrium by an extensional flow, however, their ability to relax topological constraints is suppressed. As the flow strength increases above the critical coil-stretch transition, constraints formed at equilibrium or in the startup of flow generally become relaxed only when one of the chains becomes completely stretched. At high flow strengths $Wi_c>3.0$, convection dominates, causing polymers to stretch more affinely with the flow. In this case, polymers which are hooked at equilibrium both stretch before the constraint affects the conformational dynamics.

Thus we observe a dependence of hooking probability on both concentration and flow strength, suggesting that the crossover to entanglement dynamics in polymer solutions depends on flow strength in addition to the more widely studied and understood molecular weight and concentration dependence. Our results are again consistent with those Hsiao et al. in that we observe end-coiled fast and slow hooking partners (Fig \ref{fig:hooks!}a). In this case, one polymer is stretched to become fully aligned in the flow while the other remains folded up to $\epsilon \approx 12$. 

We generally observe that hooks form at equilibrium before the startup of flow and at least one of the hooking partners becomes fully aligned by $\epsilon \approx 5$, releasing the topological constraint. These conformations often stretch faster than the ensemble average, particularly at high concentrations and flow rates, as seen in Fig \ref{fig:hooks!}c for $c/c^{*}=3.0, Wi_c = 3.49$. However, this behavior is not universal. Folded polymers can also meet and form transient constraints after significant strain accumulation. In Fig \ref{fig:hooks!}b at $c/c^{*}=1.0, Wi_c=2.13$, we observe two polymers that begin folded and unhooked, then meet at $\epsilon \approx 3$ and remain hooked until $\epsilon \approx 7$. We include simulation movies for the three cases shown in Fig \ref{fig:hooks!} in the Supplemental Information.

The representative examples shown here are only a small subset of the diverse range of hooking behavior observed in the complete ensembles. A more detailed investigation is forthcoming to statistically characterize interpolymer hooking and the crossover to entanglement behavior in polymer solutions under strong flows. While such a study is beyond the scope of this work, we expect higher molecular weight polymers and more concentrated solutions to clearly demonstrate the features observed here. In particular, we expect distinct subpopulations of polymer stretching to develop, and we conjecture that stress overshoots may emerge associated with polymers becoming stretched and aligned in flow upon the release of constraints. Already we begin to see these effects at the highest concentrations and flow rates simulated here (Fig \ref{fig:transient_ensavg}f $c/c^{*}=3.0, Wi_c=5.46$), although more conclusive results are required given the relatively large fluctuations.

\section{\label{sec:Conclusions}Conclusions}

In this work, we introduce an iterative conformational averaging method for Brownian dynamics simulation of semidilute polymer solutions in planar extensional flow. Building on our previous work, we have generalized the CA method to account for conformation dependent HI and the variation in the conformationally averaged Brownian noise as the stress is accumulated. With these refinements, we are able to nearly quantitatively reproduce transient and steady state polymer dynamics and rheology as well as conformational distributions. The modifications come at only a small increased computational cost compared to the original CA method, and we retain approximately an order of magnitude computational speedup relative to BD simulations without conformational averaging.

We have demonstrated the utility of the CA method by simulating polymers of length $N_b=100$ at the level of Kuhn step for a wide range of concentrations $c/c^{*}=0-3.0$ and flow rates $Wi_c = 0.01-5.0$. We find that the concentration dependence of the steady state fractional extension is almost entirely eliminated after rescaling by the strain rate by the concentration dependent longest polymer relaxation time. On the other hand, the extensional viscosity increases with concentration even after rescaling as solvent drag on the polymer increases due to the onset of HI screening. We also observe that conformational distributions in semidilute solutions are significantly broader than in the dilute case, even below the overlap concentration at $c/c^{*}=0.4$. Simulations at $c^{*}$ and $3c^{*}$ further reveal an interconversion between stretched and coiled conformations for flow rates above the coil-stretch transition. We attribute this to the `self-concentration' effect in semidilute solutions, which enhances changes in polymer drag with conformation due to HI screening. \cite{prabhakar2016influence}

We also find qualitative differences in the transient conformational distributions of semidilute solutions under startup extensional flow, consistent with single molecule experiments. \cite{hsiao2017direct} While this can be qualitatively explained using equilibrium-inspired approximations for conformation dependent HI, we conjecture that semidilute solutions may be fundamentally different due to the effects of topological interactions in flow. In particular, we observe that polymers can hook onto each other in the startup of flow and experience transient flow-induced entanglements. This is despite the fact that the solution is nominally unentangled at equilibrium due to the polymers ability to relax these constraints before they inhibit diffusion. We observe diverse populations of transient stretching pathways due to these intermolecular hooks. While the current results are not conclusive, we expect that further study may reveal qualitative changes in solution rheology and conformational distributions in the startup of planar extensional flow. We expect that flow-induced entanglement will significantly modify transient stretching, especially for non-linear polymer architectures. Indeed, single molecule studies have revealed that ring polymers in a linear semidilute background solution exhibit large fluctuations in fractional extension, which is likely due to threading of linear chains into the ring. \cite{zhou2019effect}

Finally, we note that the CA method still has several limitations. The conformationally averaged Brownian noise introduces error in the transient startup and relaxation dynamics. However, in this work we find this leads to only small quantitative changes. Additionally, our choice of resolution for the grid space average HI is arbitrary, and not optimized for both accuracy and computational speed. Nevertheless, due to the exact treatment of the HI in the near field, errors in the diffusion tensor are generally small. For dense solutions at high flow rates, however, hydrodynamic coupling can be long ranged, and a more refined treatment of our discrete approximation to the RPY tensor may be necessary. Finally, our method requires the computationally expensive matrix vector product $\textbf{D}_{ij} \bm{F}_j$, which scales as $O(N^{2})$. Here we are able to perform simulations of $N=12,000$ particles without significant computational expense, but we expect that for $N \sim 10^5$, the computational expense and memory requirements will become prohibitive.

\begin{acknowledgements}
This work was funded by the National Science Foundation under Grant No. CBET-1803757. The authors also acknowledge helpful comments on calculation of the Ewald sum from Ravi Prakash and Michael Graham.
\end{acknowledgements}

\appendix

\section{\label{sec:appendix}Ewald sum RPY tensor with KRBCs}
We account for the long-ranged hydrodynamic interaction (HI) via the Ewald summed Rotne-Prager-Yamakawa (RPY) tensor,\cite{beenakker1986ewald} developed by Beenakker, using a correction for bead overlap.\cite{jain2012optimization} This accounts for HI between nearest-image beads via a short-range real-space HI contribution, and also the infinite periodic images that contribute to the HI via a contribution calculated in reciprocal space. This requires judicious splitting of the short-range and long-range HI contributions so that their respective calculations converge rapidly. This is controlled by a parameter $\alpha$ that is related to the box size $\tilde{L}$ and is typically chosen to be $\alpha = 6/\tilde{L}$. The Ewald sum is then written as a sum of three contributions:
\begin{equation}
	\label{Ewald1}
    \tilde{\textbf{D}}_{ij} = \tilde{\textbf{D}}_{ij}^{self} + \tilde{\textbf{D}}_{ij}^{real} + \tilde{\textbf{D}}_{ij}^{recip}
\end{equation}
The second and third terms of this sum correspond to the real and reciprocal-space portions of the calculation between two {\it different} beads, while the first term corresponds to the self-interaction of each bead and its interaction with its own images. The terms in Equation~\ref{Ewald1} are:
\begin{equation}
	\label{Ewald2}
    \begin{aligned}
    & \tilde{\textbf{D}}_{ij}^{self} = \left(1 - \frac{6}{\sqrt{\pi}} \alpha + \frac{40}{3\sqrt{\pi}} \alpha^{3} \right) \delta_{ij} \bm{I} \\
    & \tilde{\textbf{D}}_{ij}^{real} = \sum_{\bm{n} \in Z^{3}}^{'} \bm{M}_{\alpha}^{(1)}(\tilde{\bm{r}}_{ij,\bm{n}}) \\
    & \tilde{\textbf{D}}_{ij}^{recip} = \frac{1}{\tilde{V}} \sum_{\tilde{k}_{\lambda} \ne 0} \exp(-i\tilde{\bm{k}}_{\lambda} \cdot{} 					\tilde{\bm{r}}_{ij}) \bm{M}_{\alpha}^{(2)}(\tilde{\bm{k}}_{\lambda}) \\
    \end{aligned}
\end{equation}
Here, the vector $\bm{n} = (n_{x},n_{y},n_{z})$ specifies all images including the primary image with integer components. The prime on the sum over $\bm{n}$ in the real space term indicates that the primary image $\bm{n} = (0,0,0)$ is omitted for $i=j$. The real and reciprocal portions of the diffusion matrix are functions of the vector between the images of beads $i$ and $j$, $\tilde{\bm{r}}_{ij,\bm{n}} = \tilde{\bm{r}}_{j} - \tilde{\bm{r}}_{i} + \bm{n} \cdot \bm{L}$, where $\bm{L}$ is a matrix of real-space basis vectors for a box of volume $\tilde{L}^3$. The reciprocal space basis vectors $\tilde{\bm{k}}_{\lambda}$ are formed from the reciprocal lattice of the basis vectors $\bm{L}$, and both vary with time according to the Kraynik-Reinelt boundary conditions.\cite{kraynik1992extensional,todd1998nonequilibrium}. The components are related to the basis vectors by
\begin{equation}
    \begin{aligned}
    \tilde{k}_1 = \frac{2 \pi \tilde{L}}{\textrm{det}(\textbf{L})} (n_1 L_{2y} - n_2 L_{1y}) \\
    \tilde{k}_2 = \frac{2 \pi \tilde{L}}{\textrm{det}(\textbf{L})} (-n_1 L_{2x} + n_2 L_{1x}) \\
    \tilde{k}_3 = \frac{2 \pi}{\tilde{L}} n_3
    \end{aligned}
\end{equation}
where $\textrm{det}(\textbf{L})$ is the determinant of the basis vectors matrix.
$\bm{M}_{\alpha}^{(1)}$ is a $3\times 3$ matrix:
\begin{widetext}
\begin{equation}
	\label{Ewald5}
	\bm{M}_{\alpha}^{(1)}(\bm{\tilde{r}}) = \left[C_{1} \textrm{erfc}(\alpha \tilde{r}) + C_{2} \frac{\exp(-\alpha^{2} \tilde{r}^{2})}{\sqrt{\pi}} \right] \bm{I} + \left[ C_{3} \textrm{erfc}(\alpha \tilde{r}) + C_{4} \frac{\exp(-\alpha^{2} \tilde{r}^{2})}{\sqrt{\pi}} \right] \hat{\bm{r}} \hat{\bm{r}}
\end{equation}
\end{widetext}
The $C_i$ coefficients (distinct from the $C_i$ TEA parameters) are:
\begin{equation}
	\label{Ewald6}
    \begin{aligned}
    & C_{1} = \begin{cases}
    \frac{3}{4\tilde{r}} + \frac{1}{2\tilde{r}^{3}}, & \tilde{r} \geq 2 \\
    1 - \frac{9\tilde{r}}{32}, & \tilde{r} \leq 2 \\
    \end{cases} \\
    & C_{2} = 4\alpha^{7} \tilde{r}^{4} + 3\alpha^{3}\tilde{r}^{2} - 20\alpha^{5} \tilde{r}^{2} - \frac{9}{2}\alpha + 14\alpha^{3} + \frac{\alpha}{\tilde{r}^{2}} \\
    & C_{3} = \begin{cases}
    \frac{3}{4\tilde{r}} - \frac{3}{2\tilde{r}^{3}}, & \tilde{r} \geq 2 \\
    \frac{3\tilde{r}}{32}, & \tilde{r} \leq 2 \\
    \end{cases} \\
    & C_{4} = -4\alpha^{7}\tilde{r}^{4} - 3\alpha^{3}\tilde{r}^{2} + 16\alpha^{5}\tilde{r}^{2} + \frac{3}{2}\alpha - 2\alpha^{3} - \frac{3\alpha}{\tilde{r}^{2}}
    \end{aligned}
\end{equation}
 The $3 \times 3$ matrix $\bm{M}_{\alpha}^{(2)}$ is given as
\begin{equation}
	\label{Ewald7}
    \bm{M}_{\alpha}^{(2)} = m_{\alpha}^{(2)}\left(\bm{I} - \frac{\tilde{\bm{k}}_{\lambda}\tilde{\bm{k}}_{\lambda}}{\tilde{k}^{2}}\right)
\end{equation}
where $\tilde{k} = |\bm{\tilde{k}}|$ and $m_{\alpha}^{(2)}$ is given as
\begin{equation}
	\label{Ewald8}
    m_{\alpha}^{(2)} = \left(1 - \frac{\tilde{k}^{2}}{3}\right)\left(1 + \frac{\tilde{k}^{2}}{4\alpha^{2}} + \frac{\tilde{k}^{4}}{8\alpha^{4}}\right)\frac{6\pi}{\tilde{k}^{2}} \exp \left(\frac{-\tilde{k}^{2}}{4\alpha^{2}} \right)
\end{equation}
The Ewald sum is optimized following the procedure of Jain et. al. \cite{jain2012optimization} to minimize computational time while ensuring the sum has converged to an error of approximately $10^{-4}$.

\bibliography{main}

\end{document}